\documentstyle[preprint,aps,prd]{revtex}

\begin{document}
\draft
\title{Direct $\zeta$-function approach and renormalization of one-loop stress
tensors in curved spacetimes}
\author{Valter Moretti\footnote{Electronic address: \sl
moretti@science.unitn.it}}
\address{European Centre for Theoretical Nuclear Physics and Related Areas\\
Villa Tambosi, Strada delle  Tabarelle 286
I-38050 Villazzano (TN), Italy\\
and\\
Dipartimento di Fisica, Universit\`a di Trento, \protect\\
and Istituto Nazionale di Fisica Nucleare,\protect\\
Gruppo Collegato di Trento \protect\\ I-38050 Povo (TN), Italy}

\date{Revised in September 1997}
\preprint{Preprint UTF - 399}

\maketitle

\begin{abstract}

{\small
 A method which uses a generalized tensorial $\zeta$-function to compute the
renormalized  stress tensor of a quantum field propagating 
in a (static) curved  background is presented. The method does not use 
point-splitting procedures or off-diagonal $\zeta$ functions but employs 
an analytic continuation of a generalized $\zeta$-function. The starting 
point of the method is the direct computation of the functional derivatives
 of the Euclidean one-loop effective action with respect to the background 
metric. It is proven that the method, when available, gives rise to a 
conserved stress tensor and, in the case of a massless conformally coupled 
field, produces the conformal anomaly formula  directly. Moreover, it is 
proven that the obtained stress tensor agrees with statistical  mechanics 
in the case of a finite-temperature theory.
The renormalization procedure is controlled
 by the structure of the poles of the
stress-tensor $\zeta$ function. The infinite renormalization  is automatic 
due to a ``magic'' cancellation of two poles. The remaining finite
renormalization involves locally geometrical terms arising by a certain
residue. Such terms are also conserved and thus represent just a finite 
renormalization of the geometric part of the Einstein gravitation equations 
(customary generalized through high-order curvature terms). The 
method is checked in several particular cases finding a perfect agreement with
other approaches.
First the method is checked in the case of a conformally coupled
massless field in the  static Einstein universe where all hypotheses initially
requested by the method hold true. Secondly, dropping the hypothesis of a 
closed manifold, the method is checked in the open static Einstein universe.
Finally, the method is checked for a massless scalar field  in the presence 
of a conical singularity in the Euclidean manifold (i.e. Rindler 
spacetimes/large mass black hole manifold/cosmic string manifold). \\
Concerning  the last case in particular,
the method is proven to 
give rise to the stress tensor already got by the point-splitting approach 
for every coupling with the curvature regardless of the presence of the 
singular curvature.
In the last case, comments  on the measure employed in the path integral, the
use of the optical manifold and  the different approaches to renormalize 
the Hamiltonian are made.}

\end{abstract}

\pacs{04.62.+v }
\narrowtext

\section{INTRODUCTION}

As is well-known, the stress tensor of a matter field 
in a curved spacetime  is obtained by computing the functional 
derivative of the matter field action with respect to the background metric. 
That is also the stress tensor which appears as a  gravitational source
into  Einstein`s equations. 
Trying to
generalize the theory by including quantum aspects of the matter 
fields at least,
one should consider the quantum averaged values of the same stress
tensor (considered as an operator) as a gravitational source
(see for example \cite{wald}).

As first proposed by Schwinger \cite{schwinger}, dealing with quantum 
(quasifree) field theory in curved background,  the quantum averaged 
 stress tensor is computed by executing  metric functional derivatives 
of  the one-loop effective
action. Then, the effective action  takes account of the quantum state of the 
matter fields \cite{bd}. 
In fact, considering the averaged stress tensor as gravitational source
 is the first step in order to perform a semiclassical approach to the 
quantum gravity \cite{wald,bd}. \\
One can get the averaged stress tensor also in thermal quantum
 states dealing with an opportune  Euclidean time-periodic 
 continuation of the theory
and the corresponding Euclidean effective action,
 when the 
Lorentzian manifold is static (i.e., the time of the 
considered and analytically continued coordinates defines
a time-like Killing vector normal to the surfaces at constant time).
In this case, the 
vanishing temperature limit 
should reproduce the nonthermal stress tensor referred
to the vacuum state related  to the time-like Killing 
vector\footnote{Obviously, 
 one has to eventually continue the Euclidean stress
 tensor into the real time in order to get the physical stress tensor.}. \\

The computation of the one-loop {\em regularized} and {\em renormalized} 
quantum Euclidean effective action can be performed
employing the $\zeta$-function  procedure 
\cite{hawking,bd,zerbinil} that we
 shall summarize in the following. \\
One starts with the identity
which defines the (Euclidean) effective action: 
\begin{eqnarray}
S_{\mbox{eff}}[\phi,g] :=
\ln \int {\cal D}\phi\: e^{S[\phi]} = 
-\frac{1}{2} \ln \det \left[A/\mu^2 \right]  \nonumber,
\end{eqnarray}
where $S$ is the Euclidean action of the matter field $\phi$ which we can
 suppose, for sake of simplicity,  a real scalar field 
(the approach also deals with much more complicate cases). 
The space-configuration  measure which appears in the functional integral
is that well-known \cite{hawking,fujikawa,toms}
\begin{eqnarray}
{\cal D}\phi = \prod_x \: \left\{  g(x)^{1/4} d\phi(x)    \right\} 
\label{measure}
\end{eqnarray}
and the action is built up as
\begin{eqnarray}
S[\phi] = S_A[\phi]
:= -\frac{1}{2} \int_{\cal M} d^4 x \:\sqrt{g(x)}\: \phi(x) A\phi(x).
\end{eqnarray}
where $A$ is an elliptic differential second-order selfadjoint
 operator positive defined on the Euclidean manifold
${\cal M}$. In a thermal theory with a temperature $T$, this manifold is 
periodic in the Euclidean time  being $\beta =1/T$ the period.
$\mu$ is a scale parameter necessary from dimensional considerations.
This parameter may remain in the final results and thus can be reabsorbed
into the renormalized gravitational constant as well as other 
physically measurable parameters involved in (generalized) Einstein's
 equations. This is a part of 
the programme of the semiclassical quantum gravity
approach \cite{wald,bd}.\\
We can suppose that the manifold above 
 is closed (namely compact without boundary)  
in order to have a discrete spectrum with proper eigenvectors
of $A$ and do not consider boundary conditions. Anyhow, the method 
can be generalized   for the nonclosed case (e.g. an infinite volume
 or presence of  boundaries)  by 
considering continuous spectra and boundary conditions \cite{report,zerbinil}. 
 We can compute the previous determinant in the framework of the
 {\em local} $\zeta$ function regularization \cite{hawking} by (the reason 
of that generalized definition will be clear shortly):
\begin{eqnarray}
\ln \det [ A/\mu^2] = 
-\frac{d}{ds}|_{s=0}
\zeta(s|A/\mu^2) =
-\frac{d}{ds}|_{s=0}
\zeta(s|A) -2\zeta(0|A)\ln\mu 
\label{def}.
\end{eqnarray}
The $\zeta$ function can be obtained by integrating
 the {\em local} $\zeta$ function:
\begin{eqnarray}
\zeta(s|A) = \int d^4x\sqrt{g} \:\zeta(s,x|A), \label{int}
\end{eqnarray}
where, $\phi_n(x)$ being a
normalized eigenvector
of $A$ with eigenvalue $\lambda_n$
\begin{eqnarray}
\zeta(s,x|A) = \sum_n \lambda_n^{-s} \phi_n(x)\phi^{*}_n(x) \label{locale}.
\end{eqnarray}
The expression above is the so called spectral representation of the
local $\zeta$ function.
Equivalently
\begin{eqnarray}
\zeta(s|A) = \sum_n \lambda_n^{-s} \label{locale'}.
\end{eqnarray}
These identities
 have to be understood in the sense of the analytic continuation
  of the right hand sides to values of $s$
 by which the series do not converge.
The series above converges whenever
 $ \mbox{Re}$ $s > 2 $ defining  analytic  functions which can be extended
into a meromorphic function 
defined on the  complex $s$ plane except  for two simple poles on the
real axis, at $s=1$ and $s=2$.
 We refer to \cite{zerbinil} and  references therein
 for a complete report in the
general case.\\
The main reason to define the determinant of $A$ like in (\ref{def})
is that, in the finite dimensional case,
 this coincide with the usual definition.
 One can obtain this directly through (\ref{locale'}) which reduces to
an ordinary summation in the finite dimensional case  $A$ being 
an usual matrix.\\

The $\zeta$ function approach provides us with a good theoretical 
definition of the
 determinant of an operator. Moreover, as far as the 
 quantum field theory in a curved
background is concerned, the $\zeta$-function approach
has been proven to produce the right interpretation of the functional integral
 and  the one-loop renormalized effective action whatever
 someone was  able to perform the  previously cited analytical continuation
\cite{bd,zerbinil}. Furthermore, on the theoretical ground, this
approach led to very satisfactory results.
In particular, the {\em renormalization}
 procedure\footnote{We remind the reader that the local  averaged 
quantities as 
the stress tensor  or the effective Lagrangian 
are affected by divergences also in quantum
field theory in a curved background.} hidden  in the 
$\zeta$-regularization procedure 
 seems to be the  correct one in the sense that it  
agrees with all  physical requirements and 
with  different procedures (e.g., dimensional regularization, point-splitting
method \cite{bd}). 
The important difference from the other renormalization techniques 
is that the $\zeta$-function approach leads naturally to finite quantities 
without any ``by hand'' subtraction of infinite quantities, also 
 maintaining
possible terms arising from any {\em finite} renormalization. 
Finally, it is worthwhile stressing that $\zeta$-function approaches are 
currently employed in dealing with black-hole entropy physics, in particular
to obtain quantum correction to the Beckenstein-Hawking entropy
(e.g. see \cite{cvz1,cvz2,ielmo,moiel}).\\

In principle, the Euclidean (quantum) stress tensor\footnote{When
 it is not 
specified  otherwise, it is understood
that  we  are employing
the {\em Euclidean} metric, namely the signature of the metric tensor $g_{ab}$ 
is $(1,1,1,1)$ throughout this paper.}
can be carried out from the one-loop effective action employing the usual
definition\footnote{This is the definition of the {\em Euclidean} stress 
tensor when the 
classical {\em Euclidean} 
action 
is {\em negative} definite \cite{hawking}. We adopt such a 
convention throughout this paper.}
\begin{eqnarray}
<T_{ab}(x)> = -2 g(x)^{-1/2} \:\frac{\delta 
S_{\scriptsize \mbox{eff}}[\phi,g]}{\delta 
g^{ab}(x)}. \label{tensor}
\end{eqnarray}
The Lorentzian stress tensor is then obtained by the Euclidean one
re-continuing analytically  the latter into the Lorentzian section of the 
manifold.\\
However, it is not so simple to perform the  functional derivative
in the formula written above,
 employing the $\zeta$-function regularized effective action,
 because the local $\zeta$
function is not explicitly expressed in terms of the metric. 
In general, considering all  known methods to regularize the stress tensor,
 barring (very important) theoretical consideration \cite{bd},
 it is not so simple to use the formula above at all\footnote{N.D. Birrell and
P.C.W. Davies, on page 190 of  their fundamental book \cite{bd}, wrote 
 (Birrell Davies' $W_{\scriptsize\mbox{ren}}$ is our
 $S_{\scriptsize \mbox{eff}}$):\\
``(...) {\em 
in a practical calculation it is not possible to follow this route. 
This is because in order to carry out the functional differentiation
of  $W_{\scriptsize\mbox{ren}}$} 
 {\em  with respect
to $g_{\mu\nu}$} (...), {\em it is generally necessary to know  
 $W_{\scriptsize\mbox{ren}}$ 
 for all geometries
 $g_{\mu\nu}$. This is impossibly difficult.}''}.
Other, more indirect, procedures have been found in order compute the
stress tensor, e.g. the so-called  ``point-splitting'' approach \cite{bd}
or mixed  procedures which involves  point-splitting-like methods  and 
off-diagonal local $\zeta$ functions  \cite{cvz,iellici}.\\

This paper is devoted to propose a  generalization of
 the {\em local} $\zeta$-function
approach in order to use  (\ref{tensor})
{\em directly}\footnote{A similar
attempt appeared in \cite{hawking}, but the way followed there
was quite different w.r.t. our approach because, there, 
 the heat kernel representation of the $\zeta$ function  rather than the
 $\zeta$ function expressed
in terms of eigenvalues was considered and no general theory was presented.   
An important  recent work \cite{Hu} uses  the heat-kernel representation
and further nonlocal regularization procedures to compute the 
stress tensor fluctuations in curved spacetimes.}.
We shall perform all proofs considering stationary Lorentzian 
manifolds with closed (i.e. compact without boundary) Euclidean sections.
 Anyhow, we shall see, in concrete
examples, that the method works also dropping the requirement of compactness. 
We shall present a $\zeta$-function direct 
 approach which, when available, produces a conserved
stress-tensor as well as the well-known and expected conformal anomaly
in the case of a conformally coupled massless field.
Furthermore, by our approach, one can prove thermodynamical identities
usually supposed true  without any general proof in a curved spacetime.\\
Obviously, the usual concrete problem remains, one has to perform 
some analytic continuation explicitly to get the final result and
this is not  possible,  in practice, for all physically interesting cases.
At least, the formulas
we will find define an alternative procedure among those which already
exist.  Moreover, it seems that our formulation could be  interesting 
 on the theoretical ground in particular. Indeed, as we shall see in this 
work,  one can obtain the general  results above-cited  by  employing
a very little amount of calculations and a very clear procedure. \\
Anyhow, within this paper,
 we shall consider also several particular applications of the method. 
 First, we shall consider the (thermal) theory
of a conformally coupled massless scalar field 
within the closed  Einstein universe. The Euclidean related manifold 
satisfies completely our initial hypotheses of a closed manifold.
Secondly, we shall consider   the same field propagating in the open
Einstein universe. The related Euclidean manifold is not compact and
this is a first nontrivial ground where check our approach assumed by
definition. 
The third case we shall consider  is the Euclidean manifold related
both to the cosmic string manifold and Rindler space 
(which can be considered also as the manifold containing a 
very large mass black hole). 
That Euclidean manifold 
is not {\em ultrastatic} differently from the two manifolds considered above,
moreover, it
has a conical singularity which, for some aspects,
 could be 
considered as a boundary.
That singularity involves a lot of difficulties dealing with $\zeta$
function approaches to renormalize the effective action. In particular,
  stress tensor components
 built up through the local $\zeta$ function of the effective action
in the physical manifold
have been obtained 
making direct use of mechanical-statistical laws 
 or supposing 
a particular 
form of the stress tensor {\em  a priori}.
 These results disagree, at low energies,
  with those obtained by the point-splitting method
 (see Section II of \cite{moiel} and
the final discussion  of \cite{moiel} for a  discussion and references 
on these topics).
In this paper we shall see that,  concerning the stress tensor
in the conical manifold,
 it is possible to get the same results
arising also from the point-splitting approach, for every
value of the coupling parameter $\xi$ by means of our local $\zeta$
 function approach.
This result will be carried out not  depending on the mechanical-statistical
laws and without  supposing any particular form of the stress tensor
{\em a priori}.
Concerning this case in particular but also in the general case,
 we shall point out  also some remarks on the problem
of the choice  of the configuration-space measure in the path integral 
to define the partition function of the fields. 
 We shall see that
this problem is related to the renormalization procedure involved
in defining physical quantities, concerning the Hamiltonian in particular.\\ 

The paper is organized as  follows.\\
In  {\bf Section II}, we shall build up our general approach defining
the background where it should work and we shall also stress 
some features of the method as far as the involved  finite renormalization
is concerned.\\
In {\bf Section III}, we shall analyze some general features of our theory
by employing the heat kernel expansion.\\ 
In {\bf Section IV} and {\bf V},
 we shall prove that our approach, when available,
 produces a conserved
stress tensor naturally  and gives  rise to the conformal anomaly directly
 in the case of a  
conformally invariant classical action.\\
In {\bf Section VI}, we shall prove  that our approach agrees with 
the statistical mechanics  interpretation of the time periodic Euclidean
path integral. 
This result implies  some comments on the  correct use of the apparently
``wrong'' path-integral
phase-space measure (that is an old problem  reproposed recently by
several authors).\\  
In {\bf Section VII},   we  shall compute the geometrical  tensor
related to the  finite-renormalization
part of the stress tensor in the general case of a conformally invariant
scalar field in any static curved spacetime.\\
In {\bf Section VIII}  we shall consider the simplest application
of our method, namely, we shall compute the (thermal) stress tensor of
 a massless boson field in a flat-space box.\\
In {\bf Section IX},  we shall consider the (thermal) stress tensor
of a conformally coupled massless scalar field propagating 
in  closed Einstein's universe.\\
In {\bf Section X}, 
 we shall compute the (thermal) stress tensor
of a conformally coupled massless scalar field propagating 
in  closed Einstein's universe.\\
Finally, in {\bf Section XI},  we shall compute the (thermal) stress tensor of
a massless field propagating in a manifold containing a conical singularity
in the Euclidean section for every coupling with the singular curvature.
We shall report also some comments on the thermodynamics and on the 
renormalization procedure.\\
{\bf Section XII} contains a summary of the  topics dealt with in this paper.\\
{\bf Appendix} contains  proofs of some useful formulas employed 
throughout the paper.

\section{THE $\zeta$ FUNCTION OF THE STRESS TENSOR}

Let us consider the functional definition of the stress tensor appearing in
(\ref{tensor}). In that formula, employing a $\zeta$-function approach,
the effective action is defined as:
\begin{eqnarray}
S_{\scriptsize \mbox{eff}}[\phi,g] = \frac{1}{2}\:\frac{d}{ds}|_{s=0}
\zeta(s|A)+
\frac{1}{2} \:\zeta(0|A)
\:\ln(\mu^2) \label{start}
\end{eqnarray}
where
\begin{eqnarray}
\zeta(s|A) = {\sum_n}' \lambda_n^{-s} \label{start2}.
\end{eqnarray}  
The prime means that the  summation written above does not include any 
possible null 
eigenvalues \cite{hawking}.\\
As we said in {\bf Introduction}, 
the identity (\ref{start2})
 holds in the sense of the analytic continuation
when Re$s$ $<M$, 
where $M$ is a number obtained by the heat kernel expansion
depending on the operator A and the structure of the manifold
 (usually $M=2$  dealing with Euclidean $4$-manifolds)  \cite{report,zerbinil}.
We re-stress  that the spectrum of the 
operator $A$ which appears into the Euclidean action is supposed to be 
purely discrete as it happens for Hodge-de Rham Laplacian operators in closed
manifolds \cite{chavel}. 
In other physically interesting cases, one should
 deal with proper spectral measures, or consider the 
 studied manifolds as  opportune limits of closed manifolds,
and possibly, one has to take care of possible boundary conditions
in defining the selfadjointness domain of the operator A.\\
Due to the purely heuristic form of this paper, we shall not 
consider all mathematical subtleties involved in the $\zeta$-function
approach (see \cite{chavel,zerbinil} and refs therein).

Our proposal is to perform the functional derivative with respect to the
metric  directly in the right hand side of (\ref{start2}) {\em before} we
perform  the analytic continuation. This should produce another series
and another analytic function. The value at $s=0$ of the $s$ derivative of this
new $\zeta$ function should be considered as a possible regularization
of the stress tensor.\\
In practice,
 we define the $\zeta$ {\em function of the stress tensor} as 
\begin{eqnarray}
Z_{ab}(s,x|A) \:\mbox{``} := \mbox{''} \:- 2\: g(x)^{-1/2}\:
 \frac{\delta \zeta(s|A)}{\delta g^{ab}(x)} \nonumber
\end{eqnarray} 
or, more correctly, $Z_{ab}(s,x|A)$ is the analytic continuation 
in the variable
$s$ of the series
\begin{eqnarray}
-2\: g(x)^{-1/2}\:{\sum_n}' \frac{\delta \lambda_n^{-s}}{\delta g^{ab}(x)}
= 2s\: g(x)^{-1/2}
 \:{\sum_n}' \frac{\delta \lambda_n}{\delta g^{ab}(x)} \lambda_n^{-(s+1)}
\label{series},
\end{eqnarray}
supposing that this series converges for Re $s>M'$ similarly to the 
case of the simple $\zeta$ function.\\
Then, following the spirit of (\ref{tensor}) and (\ref{start}),  our
 idea is, when possible, {\em define} the
 renormalized stress tensor as
\begin{eqnarray}
<T_{ab}(x)> := \frac{1}{2}\:\frac{d}{ds}|_{s=0}
Z_{ab}(s,x|A/\mu^2) =
 \frac{1}{2}\:\frac{d}{ds}|_{s=0}
Z_{ab}(s,x|A) 
+
 \frac{1}{2} \: Z_{ab}(0,x|A)\:\ln(\mu^2) \label{x}.
\end{eqnarray}\\
Now, our aim to get a useful expression for the function $Z_{ab}(s,x|A)$. 
In {\bf Appendix} we shall prove the formula:
\begin{eqnarray}
\frac{\delta \lambda_n}{\delta g^{ab}(x)} = 
\frac{\lambda_n}{2}\: \sqrt{g(x)} \: g_{ab}(x) \: \phi_n(x)\phi_n^*(x)
- 2\frac{\delta S_A[\phi_n^*,\phi_n]}{\delta g^{ab}(x)} \label{start3},
\end{eqnarray}
where, through obvious notations, we defined
\begin{eqnarray}
 S_A[\phi_n^*,\phi_n] :=  -\frac{1}{2} \int_{\cal M} d^4x\:\sqrt{g(x)} \:
\phi_n^*(x) A \phi_n(x). \label{snn}
\end{eqnarray}
Let us further define
\begin{eqnarray}
T_{ab}[\phi^*_n,\phi_n](x) := 
- 2\: g(x)^{-1/2}\:\frac{\delta_g S_A[\phi_n^*,\phi_n]}{\delta g^{ab}(x)} 
\label{start4}.
\end{eqnarray}
This is nothing but the classical {\em real scalar field} 
stress-tensor evaluated on the $n$th mode.
Few calculations employing (\ref{start2}) and (\ref{locale})  lead us to,
for the values of $s$ where the series in the right hand side converge
\begin{eqnarray}
-\frac{g(x)^{1/2}}{2} Z_{ab}(s,x|A) =
-s {\sum_n}' \lambda_n^{-(s+1)}\: g(x)^{1/2}\:
T_{ab}[\phi^*_n,\phi_n](x) -
\frac{s}{2} g^{1/2}(x) \: g_{ab}(x) \: \zeta(s,x|A), \label{start4.5}
\end{eqnarray}
For future reference, let us define, in the sense of the analytic continuation
in $s$
\begin{eqnarray}
\zeta_{ab}(s,x|A) := {\sum_n}' \lambda_n^{-s}\:
T_{ab}[\phi^*_n,\phi_n](x) \label{zab}.
\end{eqnarray}
It is finally useful
 to explicit the form of the function $Z_{ab}(s,x|A)$
in terms of the function  $\zeta_{ab} (s+1,x|A)$ and $\zeta(s,x|A)$.
We have
\begin{eqnarray}
Z_{ab}(s,x|A)  
&=& 
 - 2\: g(x)^{-1/2}\:
\left[ 
-s g(x)^{1/2} \zeta_{ab} (s+1,x|A) -\frac{s}{2}\: g(x)^{1/2} \:g_{ab}(x)
\zeta(s,x|A)  
\right]  \nonumber\\ 
&=& 2s\: \zeta_{ab} (s+1,x|A)
+ s  \:g_{ab}(x)
\zeta(s,x|A)  
\label{Zab}.
 \end{eqnarray} 
We stress that the functions $\zeta$ which appear in the formula above
are the analytic continuations of the corresponding series.\\
An important technical  comment is in order.
We are considering theories in which the 
$\zeta$-function approach is available in order to regularize the effective
action (Lagrangian). In such a situation the following two conditions
have to hold true: $\zeta(0,x|A)$ and $\zeta'(0,x|A)$ 
(where $'$ indicates
the $s$ derivative) must be {\em finite}. By consequence, the limits of
$s\:\zeta'(s,x|A)$ and $s\:\zeta(s,x|A)$  as $s\rightarrow 0$ have to 
vanish. 
The final result which arises performing the derivative in (\ref{x}),
taking account of the previous remark, reads 
\begin{eqnarray}
<T_{ab}(x)> &=& \left\{ 
\zeta_{ab}(s+1,x|A) + \frac{1}{2}\: g_{ab}(x) \: \zeta(s,x|A) \right.
 +\nonumber
\\
 & &\left. +  s\left[ \zeta'_{ab}(s+1,x|A) + \ln (\mu^2) \zeta_{ab}(s+1,x|A)
 \right]
\right\}_{s=0} \label{general}
\end{eqnarray}
We shall define a ``super $\zeta$-regular theory'' as a
 QFT on a (Euclidean)  manifold which can be 
regularized
by the local $\zeta$-function approach as far as the 
one-loop action and
the stress tensor  are concerned and, in particular, producing  a $x$-smooth
function
$\zeta_{ab}(s,x|A)$  which can be analytically continued from
values of $s$ where the corresponding series converges to a
 neighborhood of $s=1$ including this point.
Thus, in the case 
of a super $\zeta$-regular theory, (\ref{general}) reads more simply
\begin{eqnarray}
<T_{ab}(x)> = \zeta_{ab}(1,x|A) + \frac{1}{2}\: g_{ab}(x) \: \zeta(0,x|A)
\label{regular}.
\end{eqnarray}
Note that the stress tensor of a super $\zeta$-regular theory 
is independent of the scale $\mu$.
The price one has to pay in order to preserve the $\mu$ 
dependence is the presence of 
 a divergence  in the first term in the right hand side
of (\ref{general}). We shall come back to this point shortly.\\
The second term in the right-hand side of the equation above is quite
surprising at first sight. This is because the classical stress tensor
(evaluated on the modes) is related only with the first term in the
right-hand side. Anyhow, as we shall see later,  the unexpected terms
in (\ref{regular}) and (\ref{general}) are 
necessary in order to produce a conserved stress tensor and  give raise to the
conformal anomaly formula.
 In particular, notice that the  classical stress tensor
evaluated on the modes cannot be conserved because the modes do not satisfy
the (Euclidean) motion equations (barring  null modes).\\

In general, dealing with physical theories in four dimensional manifolds,
 we expect that the function $\zeta_{ab}(s+1,x|A)$ may take a singularity
 in $s=0$  for two reasons at least. First of all $\zeta_{ab}(s,x|A)$
is related to $\zeta(s,x|A)$ which, dealing with four dimensional manifolds,
carries a possible pole as $s\rightarrow 1$ and we
expect that $x$ derivatives do not change this fact (this will
be more clear employing the heat kernel expansion as we shall see in the 
following). A more physical reason is the following one. As is well known,
the matter-field action when renormalized through any procedure, also
different from $\zeta$-function approach (see \cite{bd}),  results to be
affected by an ambiguous part  containing an arbitrary  
 scale parameter. That role is played by  $\mu$ in the $\zeta$-function
 approach. This is a  {\em finite} 
relic of the {\em infinite} subtraction procedure. These relic terms depend
 on the geometry locally.
For this reason they can be also thought like parts of the gravitational
action \cite{bd}. In fact, it has been proven  that their only role 
is to  renormalize
 the coupling constants 
of Einstein-Hilbert's gravitational action opportunely
 generalized in order to contain also
high order terms in the curvatures \cite{bd}
\footnote{Obviously, as for the flat-space renormalization procedures,
 all  measured  physical quantities (e.g. dressed coupling constants)
are finally independent of the parameter $\mu$. See \cite{bd} for a whole
discussion.}.
\\
We have to expect that similar scale-dependent terms also appears
 in the renormalized stress
tensor. This is because
 they have to renormalize the same coupling constants which also appear
 in the geometrical
part of  (generalized) Einstein's equations of the gravity \cite{bd}.
In this sense, dealing with the stress tensor renormalization, 
the arbitrary scale $\mu$ in (\ref{general}) should play the same job
which it does as far as the $\zeta$-regularization of the effective
action is concerned.
It is worthwhile stressing that such a result 
is also allowed in Wald's axiomatic approach
to characterize  physically
 possible renormalization procedures of the stress tensor in curved spacetime
\cite{wald}. Indeed, Wald's theorem proves that a geometric 
 ambiguity remains also after one imposed strong requirements 
on the renormalization procedure. 
Such an ambiguity can be considered as an ambiguity of the coupling constants
appearing in the  geometric part of (generalized) 
Einstein's equations.
\\
Following this insight we are led to assume that, more generally than in the
 case of a super $\zeta$-regular theory, when our approach is 
available 
\begin{eqnarray}
\lim_{s\rightarrow 0} \: s\: \zeta_{ab}(s+1,x|A) =
G_{ab}(x|A) \:\:\:\:  (\mbox{finite quantity}) \label{finite}.
\end{eqnarray}
This is the only possibility in order to maintain the parameter $\mu$ into
the final renormalized stress tensor in (\ref{general}).
Our assumption implies that the function $\zeta_{ab}(s+1,x|A)$ has a simple
pole at $s=0$. \\
We shall define a ``$\zeta$-regular
theory'' as a quantum field theory on a curved
spacetime which can be regularized through the local $\zeta$ function approach
as far as the effective action is concerned, and produces a $x-$smooth 
$\zeta_{ab}(s,x|A)$  which can be analytically continued  from values 
of $s$ where the corresponding series converges to a neighborhood of $s=1$,
except for the point $s=1$ which is a simple pole.\\
 
{\em A priori}, 
in the case of a $\zeta$-regular theory, the definition 
(\ref{x})-(\ref{general})
of the renormalized stress tensor can be employed provided 
the infinite terms arising from the poles in the first and third term in the
right hand side of (\ref{general}) are  discarded.\\
Actually a magic fact happens, those two divergences cancel out
 each other and the
function $Z_{ab}(s,x|A)$ results to be 
analytic also at $s=0$ where $\zeta_{ab}(s+1,x|A)$
has a pole! 
Indeed, taking into account that the singularity in $\zeta_{ab}(s+1,x|A)$ 
is a simple pole, a trivial calculation proves that 
the structure of these singularities as $s\rightarrow 0$ 
are respectively
\begin{eqnarray}
\zeta_{ab}(s+1,x|A) 
\sim \: \frac{G_{ab}(x|A)}{s} \label{first},
\end{eqnarray}
and
\begin{eqnarray}
 s \: \zeta'_{ab}(s+1,x|A) 
\sim \: -\frac{G_{ab}(x|A)}{s} \label{second}.
\end{eqnarray}
Substituting in (\ref{general}), 
 we see that  those divergences cancel out each other.
We stress that the function $G_{ab}(x|A)$ remains into the finite 
renormalization term containing the scale $\mu$ in (\ref{general}).\\
The difference between super $\zeta$-regular
 theories and $\zeta$-regular theories concerns  only the presence of the 
scale $\mu$ in the final stress tensor.\\
As a further remark we stress that the function $G_{ab}(x|A)$ which appears 
in (\ref{finite})
 as well as in  (\ref{first})  and  (\ref{second})  contains
the  whole information about both the {\em (scale dependent) finite}
and {\em infinite}
 renormalization of the stress tensor.\\

In the next sections we shall prove the important identity which holds in case
of $\zeta$-regular theories 
\begin{eqnarray}
\nabla^a \:G_{ab}(x|A) = 0 \label{conservation}.  
\end{eqnarray}
We expect that  the term $G_{ab}(x|A)$ is built up through the local geometry
of the manifolds.  This is a consequence of the fact that the 
function $G_{ab}(x|A)$ can be carried out employing  the 
heat kernel expansion coefficients as we shall see in the next section.
All that 
means that we can consider $G_{ab}(x|A)$ as a correction to the geometrical
term in (generalized) Einstein's equations of the gravity.
This is in perfect agreement with Wald's theorem \cite{wald}.\\

We have dealt  with a real scalar field in a closed 
manifold only.
 Anyhow, reminding of the  general success of the $\zeta$-function 
approach to regularize the effective action,  
we expect that our method 
can be used to regularize the stress tensor in
more general situations,  simply passing, when necessary, to consider
(charged) spinorial modes\footnote{The case of non integer spin 
could be more complicated. In the case of gauge fields, it is convenient
to employ the Hodge-de Rham formalism and one has to
take the ghost contribution to the stress tensor into account.} 
and continuous spectral measures
in (\ref{zab}) and (\ref{regular}).  Conversely, 
the presence of boundaries could involve further problems.
The examples we shall consider in {\bf Sections VIII, XI, XII} deal with
some possible generalizations. 

\section{HEAT-KERNEL EXPANSION ANALYSIS}

In this section we shall consider, on a general ground, the behavior 
of the function $Z_{ab}(s,x|A)$ near the point $s=0$  in the case of
a real scalar field whose action is 
\begin{eqnarray}
 S =  -\frac{1}{2} \int d^4 \: \sqrt{g}\:
\left( \nabla_a \phi \nabla^{a} \phi + m^2 \phi^2 + \xi R \phi^2
\right),
\label{actionend}
\end{eqnarray}
By employing the heat-kernel expansion 
we shall see that such a theory define  a
$\zeta$-regular theory  (possibly also super $\zeta$-regular).
 We shall be also able to relate the residue $G_{ab}(s,x|A)$ to the heat-kernel
coefficients. \\

The operator which correspond to the action above is
\begin{eqnarray}
A = -\Delta + m^2 + \xi R \label{operatorend}
\end{eqnarray}
and  the  corresponding stress tensor reads
\begin{eqnarray}
T_{ab}(x) =\nonumber
\end{eqnarray}
\begin{eqnarray}
\nabla_a \phi \nabla_b\phi - \frac{1}{2}g_{ab}(x)\left(
\nabla_c \phi \nabla^{c} \phi + m^2 \phi^2\right)  
+ \xi \left[\left( R_{ab}- \frac{1}{2}g_{ab} R \right) \phi^2
+g_{ab} \nabla_c \nabla^{c} \phi^2 - \nabla_a\nabla_b \phi^2 \right]
\label{stressend}.
\end{eqnarray}
A few calculations lead to the stress tensor evaluated on the modes
\begin{eqnarray}
T_{ab}[\phi^*_n\phi_n](x) &=&
\frac{1}{2}\left(\nabla_a \phi^*_n \nabla_b\phi_n +
\nabla_b \phi^*_n \nabla_a\phi_n\right)
-\xi \nabla_a\nabla_b |\phi_n|^2 \nonumber\\
& &+ \left(\xi- \frac{1}{4} \right)
g_{ab} \Delta |\phi_n|^2 + \xi R_{ab} |\phi_n|^2 - \frac{1}{2} g_{ab}
\lambda_n
|\phi_n|^2
\label{modstressend}.
\end{eqnarray}
We are able to write down the function $\zeta_{ab}(s,x)$ in the general case
considered above. Employing the definition (\ref{zab}) we find
\begin{eqnarray}
\zeta_{ab}(s,x|A) &=& \frac{1}{2}{\sum_n}' \:\lambda_n^{-s}
\left(\nabla_a \phi^*_n \nabla_b\phi_n +\nabla_b \phi^*_n \nabla_a\phi_n
\right) \nonumber\\
& & + \left[-\xi \nabla_a\nabla_b + \left(\xi- \frac{1}{4} \right)
g_{ab} \Delta  + \xi
R_{ab} \right] \zeta(s,x|A) \nonumber \\
& & - \frac{1}{2}
g_{ab} \zeta(s-1,x|A) \label{zabend}.
\end{eqnarray}
For future reference, it is convenient to define also
\begin{eqnarray}
\bar{\zeta}_{ab}(s,x|A) := \frac{1}{2}{\sum_n}' \:\lambda_n^{-s}
\left(\nabla_a \phi^*_n \nabla_b\phi_n +\nabla_b \phi^*_n \nabla_a\phi_n
\right), \label{barzabend}
\end{eqnarray}
where we suppose to continue 
the series above analytically 
as far as possible in the complex $s$ plane.\\
We want to study the behavior of the function $\zeta_{ab}(s+1,x|A)$
 and hence $Z_{ab}(s,x|A)$ near the possible singularity at $s=0$
and, more generally,
we want to study the meromorphic structure of these functions.\\
Let us consider the  off-diagonal heat-kernel asymptotic  expansion
 \cite{fulling}
in four dimension,
which holds asymptotically for $t\rightarrow 0$ and $x$ near $y$
(in a convex  normal neighborhood)
in closed Euclidean manifolds
\begin{eqnarray}
H(t,x,y|A) \sim (4\pi t)^{-2}\: e^{-\sigma(x,y)/2t} \sum_{j=0}^{+\infty}
a_j(x,y|A)\: t^j \label{heatkernel}.
\end{eqnarray}
$\sigma(x,y)$ is half the square of the geodesical distance
from the point $x$ to the point $y$. The heat kernel 
$H(t,x,y)$ decays very speedly as $t\rightarrow +\infty$, the only
singularities come out from its behavior near $t=0$ when $x=y$.\\
The relation between the heat-kernel expansion and the $\zeta$-function theory
(\cite{zerbinil}) is that the  heat kernel $H(t,x,y|A)$ satisfies
\begin{eqnarray}
\zeta(s,x,y|A) = \frac{1}{\Gamma(s)} \int_0^{+\infty} dt\: t^{s-1}
H(t,x,y|A)  \label{relation},
\end{eqnarray}
where, for Re $s$ sufficiently large 
\begin{eqnarray}
\zeta(s,x,y|A) = {\sum_n}' \lambda_n^{-s} \: \phi^*_n(x)\phi_n(y). 
\label{relation2}
\end{eqnarray}
We can decompose the integration above into two parts as
\begin{eqnarray}
\zeta(s,x,y|A) = \frac{1}{\Gamma(s)} \int_0^{1} dt\: t^{s-1}
H(t,x,y|A)  +
 \frac{1}{\Gamma(s)} \int_1^{+\infty} dt\: t^{s-1}
H(t,x,y|A) 
\label{relation3}.
\end{eqnarray}
The true 
difference from left hand side and right hand side of (\ref{heatkernel})
is a function regular as $t\rightarrow  0$.
Taking into account that fact, one can insert the expansion in 
(\ref{heatkernel}) 
into the first integral in the right hand side of (\ref{relation3}),
obtaining
\begin{eqnarray}
\zeta(s,x,y|A) =  \frac{1}{\Gamma(s)} \sum_{j=0}^{+\infty} 
a_j(x,y|A)\: \int_0^{1} dt\: t^{s-3+j} e^{-\sigma(x,y)/2t} + h(s,x,y|A)
\label{relation4},
\end{eqnarray}
where $h(s,x,y)$ is a unknown $x,y$-smooth and  $s$-analytic  function.
This relation is the starting point of our considerations.\\
As  general remarks we stress the following two fact.  \\
First, the coefficients
$a_j(x,y|A)$ expressed in {\em Riemannian coordinates} centered in $x$
(\cite{bd}) are polinomials in $x-y$  whose coefficients
are algebraic combinations of curvature tensors evaluated at the  point
$x$. Thus the limit as $x \rightarrow y$
of  quantities as $a_j(x,y|A)$, $\nabla^{(x)}_a a_j(x,y|A)$,
$\nabla^{(x)}_a \nabla^{(y)}_b
 a_j(x,y|A)$ and so on  we shall consider shortly,
produces algebraic combinations of (covariant derivatives of) curvature
tensors evaluated at the same point $x$. \\
Secondly, there exists a recursive procedure which permits to get the
coefficients
 $a_j(x,y|A)$ and their covariant derivatives evaluated in the limit
of coincidence of arguments, when are known the coefficients $a_i(x,y|A)$,
their derivative  for $0 \leq i < j$ and the
covariant derivatives of the function $\sigma(x,y|A)$, everything
evaluated in the argument coincidence  limit.
Such a procedure can be obtained by a simple generalization of a 
similar procedure (which does not consider covariant
derivatives) presented in \cite{fulling}. \\

Let us evaluate the pole structure of the function $\bar\zeta_{ab}(s,x|A)$
employing (\ref{relation4}) and the following known identities 
\cite{bd,fulling}
\begin{eqnarray}
\nabla^{(x)}_a \nabla^{(y)}_b \sigma(x,y)|_{x=y} &=& -g_{ab(x)}\nonumber,\\
\nabla^{(x)}_a \sigma(x,y)|_{x=y} &=& 0\nonumber,\\
\nabla^{(y)}_b \sigma(x,y)|_{x=y} &=& 0\nonumber.
\end{eqnarray}
 By taking the opportune derivatives in (\ref{relation4})
and posing $y=x$ finally, we find
\begin{eqnarray}
\bar\zeta_{ab}(s+1,x|A)  =
\bar\zeta_{ab}(s+1,x|A)_{\scriptsize \mbox{analytic}}+\nonumber
\end{eqnarray}
\begin{eqnarray}
 \frac{1}{(4\pi)^2\Gamma(s+1)}
\left[ \frac{a_{0(ab)}(x|A)}{s-1} +\frac{a_{1(ab)}(x|A)}{s}
  +\frac{1}{2}g_{ab}(x) \left(
\frac{a_{0}(x|A)}{s-2} +\frac{a_{1}(x|A)}{s-1}
+ \frac{a_{2}(x|A)}{s}
  \right)\right] \label{poles}
\end{eqnarray}
where we defined
\begin{eqnarray}
a_{j(ab)}(x|A) :=\frac{1}{2}\: \left[
 \nabla^{(x)}_a\nabla^{(y)}_b a_{j}(x,y|A)
+ \nabla^{(y)}_a\nabla^{(x)}_b a_{j}(x,y|A)\right]_{x=y} \label{def2}.
\end{eqnarray}
Notice that in the pole expansion  written above, an infinite number of
apparent  poles 
have been  canceled out by corresponding   zeros of  $(\Gamma(s+1))^{-1}$.\\
The function $\zeta(x|A)$ has the well-known similar structure
\begin{eqnarray}
\zeta(s,x|A)  &=& 
\zeta_{ab}(s,x|A)_{\scriptsize \mbox{analytic}} + \frac{1}{(4\pi)^2\Gamma(s)}
\left(\frac{a_{0}(x|A)}{s-2} +\frac{a_{1}(x|A)}{s-1}\right)  \label{polesz}.
\end{eqnarray}
Employing the results  written above 
to calculating the pole structure of the function\\
$\zeta_{ab}(s+1,x|A)$ through (\ref{zabend}), we find 
\begin{eqnarray}
(4\pi)^2 \zeta_{ab}(s+1,x|A) =
(4\pi)^2 \zeta_{ab}(s+1,x|A)_{\scriptsize \mbox{analytic}} \nonumber
\end{eqnarray}
\begin{eqnarray}
& & +
 \frac{1}{\Gamma(s+1)}
\left[ \frac{a_{0(ab)}(x|A)}{s-1} +\frac{a_{1(ab)}(x|A)}{s}
+\frac{g_{ab}(x)}{2}
 \left(
\frac{a_{0}(x|A)}{s-2} +\frac{a_{1}(x|A)}{s-1}
+ \frac{a_{2}(x|A)}{s} \right)
\right]  
 \nonumber\\
& & +   \frac{1}{\Gamma(s+1)}
\left[-\xi \nabla_a\nabla_b + \left(\xi- \frac{1}{4} \right)
g_{ab} \Delta  + \xi
R_{ab} \right] 
\left(
\frac{a_{0}(x|A)}{s-1} +\frac{a_{1}(x|A)}{s}
\right)  
 \nonumber \\
& & - 
\frac{g_{ab}(x)}{2\Gamma(s)} 
  \left(
\frac{a_{0}(x|A)}{s-2} +\frac{a_{1}(x|A)}{s-1}
\right) 
\label{polefine}.
\end{eqnarray}
We stress the presence of a simple pole for $s=0$.
The pole expansion above written proves that the considered theory
is a $\zeta$-regular theory. The theory is also a {\em super}
$\zeta$-regular theory when the residue at $s=0$ vanishes.\\
This residue is just the function $G_{ab}(x|A)$
 which reads
in terms of heat-kernel coefficients
\begin{eqnarray}
G_{ab}(x|A) 
= \nonumber
\end{eqnarray}
\begin{eqnarray}\frac{1}{(4\pi)^2}
\left\{ a_{1(ab)}(x|A) + \frac{g_{ab}(x)}{2} a_2(x|A) +
\left[-\xi \nabla_a\nabla_b + \left(\xi- \frac{1}{4} \right)
g_{ab} \Delta  + \xi
R_{ab} \right] a_1(x|A)
 \right\} \label{Gexplicit}.
\end{eqnarray}
Now,
it is obvious that $G_{ab}(x|A)$ depends on the geometry locally. In particular
it is built up by algebraic combination of curvature tensors and their
covariant derivatives. A closer scrutiny, employing the recursive procedure
to compute the heat-kernel coefficients
cited above, proves that $G_{ab}(x|A)$ contains combinations of
products of two curvature tensors at most\footnote{In particular,
in a flat space and for $m=0$ the residue above  vanishes.}. Considering that
$G_{ab}(x|A)$ is also conserved, this means that it is obtained
from an Einstein-Hilbert action improved by including quadratic terms
in the curvature tensors. 
As we said, $G_{ab}(x|A)$ is the part of the renormalized
stress tensor which can be changed by finite renormalization. 
 This agrees with all known 
different stress-tensor renormalization procedure
where one finds that, in the case of a scalar
field studied here,  the finite renormalization of the stress tensor involves
only curvature quadratic terms \cite{bd}. This agrees  with Wald's
theorem and the related comments reported in \cite{wald}, too.
We shall return on these facts later.

\section{CONSERVATION OF THE STRESS TENSOR AND $G_{ab}(x|A)$}

Let us prove that, in the case of a super $\zeta$-regular theory or
a $\zeta$-regular theory,
 the stress tensor obtained from (\ref{general}) is
conserved. By the same proof, we shall get  conservation of 
$G_{ab}(x|A)$ too.\\
Our strategy will be the following one. We shall consider the function
whose the value at $s=0$ is the renormalized stress tensor
\begin{eqnarray}
<T_{ab}(s,x)> :=\frac{1}{2}\:\frac{d}{ds}
Z_{ab}(s,x|A)
+\frac{1}{2} \: Z_{ab}(s,x|A)\:\ln(\mu^2) \label{x'} 
\end{eqnarray}
and we shall evaluate the covariant divergence for the values of $s$
in which the involved $\zeta$-function can be expanded as a series.
We shall find that this covariant divergence vanishes. 
Due to the analyticity of the considered functions, this result
can be   continued as far as  the physical value  $s=0$.

Let us consider the $\zeta$ function $Z_{ab}(s,x)$ expressed as the series
in (\ref{series})
\begin{eqnarray}
Z_{ab}(s,x) = 2s\: g(x)^{-1/2}\:
 {\sum_n}' \lambda_n^{-(s+1)}
 \frac{\delta \: \lambda_n}{\delta g^{ab}(x)}\nonumber.
\end{eqnarray}
We have
\begin{eqnarray}
\nabla^{a}  Z_{ab}(s,x)  = 
 -s {\sum_n}' \lambda_n^{-(s+1)}\:
 \nabla^{a}
\:\left[
-2 g(x)^{-1/2}\:
\frac{\delta \lambda_n}{\delta g^{ab}(x)} \right] \label{qf}.
\end{eqnarray}
Let us prove that 
\begin{eqnarray}
\nabla^{a}\:Z_{ab}(s,x) = 0 \label{qf'}
\end{eqnarray}
because
\begin{eqnarray}
\nabla^{a}  \: \left[ -2  g(x)^{-1/2}
\frac{\delta \lambda_n}{\delta g^{ab}(x)} \right] = 0 \label{fine1}.
\end{eqnarray}
Due to (\ref{qf}) and (\ref{x'}), this proves conservation of the stress
tensor by taking the limit at $s=0$.\\
A nice proof of (\ref{fine1}) deals with as it follows. Let us consider the 
new ``action'' 
\begin{eqnarray}
\Lambda_n[g,\phi^*,\phi] := 2 S[g,\phi^*,\phi] -\lambda_n \int_{\cal M}
d^4x \:\sqrt{g(x)} \: \phi^*(x) \phi(x) \nonumber.
\end{eqnarray}
This is  a diffeomorphism invariant action producing the field equations
\begin{eqnarray}
A\phi(x) = \lambda_n \phi(x)\:\:\:\:\: \mbox{and}\:\:\:\:\:
A\phi^*(x) = \lambda_n \phi^*(x)
\nonumber.
\end{eqnarray}
In particular, these equations are fulfilled by the eigenfunctions $\phi_n(x)$
 and $\phi^*_n(x)$. As well-known, due to diffeomorphism 
invariance of the action,
 one gets  conservation 
of a stress tensor $T_{n\:\:ab}(x) $  evaluated  on the motion solutions,
namely, {\em on the modes} $\phi_n(x)$ {\em and}
 $\phi^*_n(x)$.
Again, this stress tensor is  obtained as the functional derivative of the
action $\Lambda_n$ with respect to the metric (with the overall factor 
$-2 g(x)^{-1/2}$). A little computation and (\ref{start3}) get just
 \begin{eqnarray}
T_{n\:\:ab}(x) 
:= -2g(x)^{-1/2} \frac{\delta_g \Lambda_n}{\delta g^{ab}(x)} =
  2g(x)^{-1/2} \frac{\delta \lambda_n}{\delta g^{ab}(x)}.
\end{eqnarray}
Conservation of the left hand side implies (\ref{fine1}) trivially.\\

An important remark, in the case of a $\zeta$-regular theory,
is finally necessary. 
 Conservation of the tensor $Z_{ab}(s,x)$ reads, employing (\ref{Zab})
\begin{eqnarray}
s \nabla^{a} \: \left\{\zeta_{ab}(s+1,x|A) +g_{ab}(x) \:\zeta(s,x|A) 
\right\} =0. \nonumber
\end{eqnarray}
We get, recalling (\ref{first})
and performing the limit as $s\rightarrow 0$ 
\begin{eqnarray}
\nabla^{a} \: G_{ab}(x|A) =0. \nonumber
\end{eqnarray}
This is nothing  but (\ref{conservation}).

\section{THE CONFORMAL ANOMALY}

Let us prove of the conformal anomaly  formula \cite{bd}
by 
employing a way similar to that in the previous section,
in the case of a super $\zeta$-regular theory or
a $\zeta$-regular theory.\\
As usually,  we have to  suppose that
the classical action $S[\phi]$ is conformally invariant. As well-known, by
performing  an infinitesimal
 local  conformal transformation on both the metric and the field,
the following equations arise
\begin{eqnarray}
g^{ab}(x)T_{ab}[\phi](x) -g(x)^{-1/2}\: \phi(x)
 \frac{\delta S}{\delta \phi(x)} =0. \nonumber
\end{eqnarray} 
This implies that classically, 
{\em working on solutions of the motion equation}, the trace
of the stress tensor vanishes. Similarly, dealing with
 the action evaluated on the modes
$\phi_n(x)$,  conformal invariance of the action lead us to\footnote{Notice
that we transform the modes employing the same transformation  of the
 field $\phi(x)$. This transformation
does not preserves the normalization of the modes but preserves the 
value of the action.}
\begin{eqnarray}
g^{ab}(x)T_{ab}[\phi^*_n\phi_n](x)
 -g(x)^{-1/2}\: \phi_n(x)
 \frac{\delta S}{\delta \phi_n(x)}
 -g(x)^{-1/2} \phi^*_n(x)
\frac{\delta S}{\delta \phi^*_n(x)}
 =0 \nonumber
\end{eqnarray} 
or equivalently
\begin{eqnarray}
g^{ab}(x)\:T_{ab}[\phi^*_n\phi_n](x) =
-\:\lambda_n\: \phi^*_n(x)
 \phi_n(x) \nonumber
\end{eqnarray} 
From this equation, employing (\ref{locale}) and (\ref{zab})
 we get
\begin{eqnarray}
g^{ab}(x)\: \zeta_{ab}(s+1,x|A) = - \zeta(s,x|A), \label{anomaly}
\end{eqnarray}
 where the involved  $\zeta$ functions can be 
defined as series.\\
Holding our hypothesis of a $\zeta$-regular theory,
 this result can be analytically continued arbitrarily close to
the physical value $s=0$. In particular, the left hand side of (\ref{anomaly})
 must be 
{\em finite} at $s=0$ because  so is the right hand side.
This seems quite
surprising because $\zeta_{ab}(s+1,x|A)$ may take a pole at $s=0$. 
We conclude that the pole 
has to disappear due to trace procedure in case of a 
conformally invariant action, namely
\begin{eqnarray}
g^{ab}(x) G_{ab}(x|A) = 0. \label{trg}
\end{eqnarray}
We shall check this fact directly later.  \\
It is worthwhile noticing
that the trace procedure, canceling out the pole in $g^{ab}(x)
\zeta_{ab}(1,x|A)$, 
gives rise to  vanishing terms 
$s\: g^{ab}(x)\zeta'_{ab}(s+1,x|A)$ and 
$s\: g^{ab}(x)\zeta_{ab}(s+1,x|A)$ when $s
\rightarrow 0$.
Finally, (\ref{general}) through (\ref{anomaly})
produces the well-known conformal anomaly
formula \cite{hawking,bd}
\begin{eqnarray}
g^{ab}(x)\:<T_{ab}(x)> = \zeta(0,x|A).  \label{conformal anomaly}
\end{eqnarray}

\section{THERMODYNAMICS AND COMMENTS ON THE 
PHASE-SPACE MEASURE OF THE PATH INTEGRAL}

In this section
\footnote{From now on, we employ
the signature $(-1,1,1,1)$ for the Lorentzian metric and 
Lorentzian quantities shall be labeled by an index $L$.}
we prove that for $\zeta$-regular theories
or  super $\zeta$-regular theories
\begin{eqnarray}
-\frac{\partial \ln Z_\beta
}{\partial\beta} \:
= -\int_\Sigma d\vec{x} \:\sqrt{-g_L}\: <T_{L\: 0}^{\:\:0}(\vec{x})>_\beta.
 \label{thermal}
\end{eqnarray}
where we have defined $\ln Z_\beta := S_{\scriptsize \mbox{eff} }$,
provided the (Euclidean and Lorentzian) 
manifold admits a
global  (Lorentzian time-like)
Killing vector arising from the Euclidean temporal coordinate with
a  period $\beta =1/T\:$. 
 $\vec{x}$ represents the spatial coordinates which belong to the 
spatial section $\Sigma$ and $g_L=-g$ 
is the determinant of the Lorentzian metric. \\
As it is clear from the notations,
we are trying to interpret $Z_\beta$ as a {\em partition function} 
 \footnote{$T$ is the ``statistical'' 
temperature, the ``local thermodynamical''
one being given by Tolman's relation $T /\sqrt{g_{00}}$
 $(= T/ \sqrt{-g_{L\:00}})$. }.\\
Notice that
all quantities which appear in the formula above do not depend 
on the Euclidean or Lorentzian time  because  the manifold is stationary
and thus no time dependence arises from the metric.
By the same reason, the time dependence in the eigenvectors 
of the motion operator is exponential and thus it cancels out 
in all involved local $\zeta$ functions.
Finally, $<T_{L\: 0}^{\:\:0}(\vec{x})> =<T_{0}^0(\vec{x})>$
 by a trivial analytic 
continuation.\\
Actually, it is not necessary to interpret $x^0$ as a time coordinate, the same
result in (\ref{thermal}) arises also when the Killing vector is associated to
the ``spatial'' coordinate $x^{i}$, provided $\beta$ were changed to $L_{i}$,
the ``spatial'' period of the manifold along the $i$th direction.
Assuming both the homogeneity along $x^0$ and $x^{i}$ we get another
expected formula trivially:
\begin{eqnarray}
-\frac{\partial \ln Z_\beta}{\partial L_i} \:
= -\frac{\beta}{L_i} 
\int_\Sigma d\vec{x} \:\sqrt{-g_L}\: <T_{L\: i}^{\:\:i}(\vec{x})>_\beta.
 \label{thermal2}
\end{eqnarray}

Before we start with the proof of (\ref{thermal}),
 some important remarks are in order.\\
In particular, let
 us consider a scalar field with an Euclidean action coupled with 
the scalar curvature,  given by
\begin{eqnarray}
S[\phi] &=& -\frac{1}{2} \int \:d^4 x\: \sqrt{g(x)} \phi(x) \:A\:\phi(x) =
\nonumber\\
& & -\frac{1}{2} \int \:d^4 x \: \sqrt{g(x)}\: \phi(x)\left[ 
-\nabla_a\nabla^{a}
+ m^2   + \xi R(x) \right] \phi(x) \label{action}
\end{eqnarray}
and let us assume explicitly that the (both Lorentzian and Euclidean) 
metric is {\em static}, namely
$g_{(L)0i}=0$ besides $\partial_{0} g_{(L)ab}(x) =0$ (but not necessarily
{\em ultrastatic}). 
In that case, in principle
\cite{hawking}, there is no problem in implementing the 
 canonical-ensemble approach to the thermodynamic and trying the
 interpretation
of the Euclidean time-periodic path integral as a  partition function
$Z_\beta$, and thus, in principle,
\begin{eqnarray}
-\beta^{-1} S_{\scriptsize \mbox{eff}} = -\beta^{-1}\ln Z_\beta \nonumber
\end{eqnarray}
could be interpreted as the free energy of the field in the
considered quantum thermal state.
The case of a stationary manifold ($g_{L\:0i}\neq 0$) involves more subtleties
also considering the analytic continuation into an Euclidean manifolds
which we shall not consider here \cite{hawking}. Anyhow, it is worthwhile
stressing that (\ref{thermal}), written in terms
of $<T^0_0>$  and $g$, holds true in the general case of a stationary
Euclidean metric $\ln Z_\beta$ being $S_{\scriptsize \mbox{eff}}$
without assuming that this define any free energy.\\
 Identities as  (\ref{thermal})  or (\ref{thermal2})
 represent a direct evidence that the definition of the
partition function as a path integral on the 
continued Euclidean  manifold, also in the case of a {\em curved} spacetime, 
 does not lead to thermodynamical inconsistencies
  in the case of a closed spatial section of the manifold at least.
We stress that $-T_0^0$ does not coincide with
the Hamiltonian density ${\cal H}$ which one  could expect
 in the right hand side of 
(\ref{thermal}). Anyhow,  the difference of these quantities is a 
spatial divergence which does not produce contributions to the spatial
integral, holding our hypothesis of a closed spatial section. Indeed, 
the case of a {\em static metric}
we have 
\begin{eqnarray}
{\cal H} =   -T^0_0  + \xi (-g_L)^{-1/2} \partial_i [(-g_L)^{1/2}
(g^{ij} \partial_j \phi^2   \:- \phi^2 w^{i}) ], 
\end{eqnarray}
  where $w^{a} = \frac{1}{2}
\nabla^{a} \ln |g_{L00}|$. 
Interpreting $<\phi^2(x)>$ as the limit of $\zeta(s,x|A)$ 
as $s\rightarrow 1$, the previous equation leads
to a natural regularization of $<\int d\vec{x}\:\sqrt{g(\vec{x})}\:
 {\cal H}(x)>$ which coincides with the corresponding 
integral of $<T^0_0(\vec{x})>$ which appears in 
the right hand side of (\ref{thermal})
 \footnote{One has to be very careful in dealing
with the limit as $s \rightarrow 1$ (I am grateful to D. Iellici 
who has focused my attention on this general problem) 
 because as previously discussed, in four dimensions,
 $\zeta(1,x|A)$ usually diverges
 as it follows from heat kernel theory \cite{zerbinil}, except for the case
of a massless field  conformally coupled to $R$ or a massive field 
with an opportune  coupling with $R$
in a curvature-constant manifold. 
Actually, one has to calculate {\em first} the spatial 
integral for $s\neq 1$ and thus all  
terms containing the integral of  the derivative of $\zeta(s,x|A)$ 
 on
$\partial \Sigma$  vanish,
{\em then} one can perform the limit as $s\rightarrow 1$
which is trivial.}.
 
The validity of
 (\ref{thermal})  and  (\ref{thermal2})  
is an indirect proof that the 
canonical measure suggested by Toms \cite{toms} 
in defining the path integral in the phase space 
\begin{eqnarray}
 \prod_x \:\left\{[g^{00}(x)]^{-1/2} \:  d\phi(x) d\Pi(x)\right\} \nonumber 
\end{eqnarray}
instead of the  apparently more ``natural''\cite{citetoms,dealwis} 
\begin{eqnarray}
\prod_x \:  \left\{d\phi(x) d\Pi(x)\right\}\nonumber
\end{eqnarray}
  can be
correctly used in defining the partition function in terms of an Euclidean
Hamiltonian path integral.
 Indeed it is Toms'  measure in the phase space which
produces, by the usual momentum integration, the configuration space
measure  (\ref{measure}) which is 
used as a starting point to the $\zeta$-function interpretation
of the configuration space path integral \cite{hawking,toms,fujikawa}.  \\
As a final comment, it is worthwhile  stressing
 that,
  already on a classical ground,
 dropping the requirement of a closed spatial section, the
Hamiltonian could not coincide with the integral of $T^0_0$ and the theory 
would be more problematic.
This could be very important in studying the quantum correction of the 
black-hole entropy, where the spatial section of the spacetime has a boundary
represented, in the Lorentzian picture, by the event horizon
\cite{FF}.\\

To conclude, let us prove the identity (\ref{thermal}). We just sketch 
the way because that is very similar to the proofs in the previous sections.
In {\bf Appendix} we shall prove the identity (where $g_0^0 = 1$)
\begin{eqnarray}
\frac{\partial \lambda_n}{\partial\beta} = -2 \int_\Sigma d\vec{x}
\sqrt{g(\vec{x})} \left\{ T_0^0[\phi^*_n\phi_n](\vec{x}) 
+ \frac{1}{2}\:g^0_0 \:\lambda_n \: \phi_n^*(\vec{x})\phi_n(\vec{x}) \right\}
\label{endend}.
\end{eqnarray}
From the expression above and employing  definitions in {\bf Section II},
 we get that, for the values of $s$ where the involved $\zeta$
functions can be expanded as series 
\begin{eqnarray}
 \frac{\partial\zeta(s|A)}{\partial\beta} &=& \int_\Sigma d\vec{x}
\: \sqrt{g(\vec{x})}\: 2s \left\{  \zeta_0^0(s+1,\vec{x}|A)
+\frac{1}{2}g^0_0\:\zeta(s,\vec{x}|A)\right\} \nonumber\\
&=&  \int_\Sigma d\vec{x}\sqrt{g(\vec{x})}\: Z^0_0(s,x|A), \nonumber
\end{eqnarray}
and thus we find
\begin{eqnarray}
 -\frac{\partial \ln Z_\beta}{\partial\beta}\: =
-\frac{\partial S_{\scriptsize \mbox{eff} }}{\partial\beta} 
=\frac{1}{2}\frac{d}{ds}|_{s=0}
\int_\Sigma d\vec{x}\sqrt{g(\vec{x})}\: Z^0_0(s,x) + 
\frac{1}{2}\ln(\mu^2) \int_\Sigma d\vec{x}\sqrt{g(\vec{x})}\: Z^0_0(0,x)
  =\nonumber 
\end{eqnarray}
\begin{eqnarray}
= -\int_\Sigma d\vec{x} \:\sqrt{g}\: <T_{0}^{0}(\vec{x})>_\beta.
\nonumber
\end{eqnarray}
That is (\ref{thermal}).
Notice that both $Z_\beta$ and $T^0_0$ may be affected by arbitrary 
 $\mu$-dependent terms.
A comparison between both sides of (\ref{thermal})
explicited in terms of $\zeta$ functions lead us to
the identity for the factors of $\ln (\mu^2)$
\begin{eqnarray}
\frac{\partial \zeta(0|A)}{\partial \beta}
= 2 \int_\Sigma d\vec{x} \:\sqrt{g}\: G^0_0(\vec{x}|A),
\end{eqnarray}
where $G_{ab}(x|A)$ is the previously introduced 
residue of $\zeta_{ab}(s+1,x|A)$ at $s=0$ (\ref{first}).

\section{Explicit computation of $G_{ab}(x|A)$ in a $\zeta$-regular theory:
the  conformally coupled case}

Let us consider the case of a massless scalar
 field conformally coupled in a generic (closed Euclidean) four
 dimensional spacetime.\\
Because a particular discussion on the form of $<T_{ab}>$ 
depends on the particular manifold we are 
dealing with, we shall consider, in the general case of
a massless conformally coupled field, only the general form of
the pole $G_{ab}(s,x|A)$
 employing the equations founds in {\bf Section III}.
We shall find that   $G_{ab}(s,x|A)$ has a vanishing trace
(and thus the conformal  anomaly formula follows as we saw previously),
it is  conserved 
and depend locally on the geometry. In particular it is quadratic
in the curvatures and can be thought as a generalization of the
geometrical term in Eintein's equations.
Moreover, we shall find that the explicit form of $G_{ab}(x|A)$ 
is just that required by other renormalization procedures.

We remind the reader the first and the second  heat kernel off-diagonal
 coefficient
in the case of a massless field. These coefficients appears in
\cite{bd}\footnote{It is very important to note that
the coefficients reported in \cite{bd} are referred to the Lorentzian
metric. The choice of the signature employed in \cite{bd} is $(1,-1,-1,-1)$
and the definition of the Riemann tensor $R^{a}_{bcd}$ takes the opposite sign
with respect the more usual  choice  \cite{fulling} which we are employing.
To pass from the Lorentzian convention in \cite{bd} to our Euclidean convention
is  sufficient  to use the two formal 
transformations $R'^{a}_{bcd} \rightarrow - R^{a}_{bcd}$,
$g'_{ab}\rightarrow -g_{ab}$, where the primed quantities are those
Lorentzian which appear in \cite{bd} and the others are our Euclidean 
quantities. The definitions of $R_{ab}$ and $R$ do not change, we have
 $R_{ab} := R^{c}_{acd}$, $R:= R^c_c$ in both formalisms.} 
\begin{eqnarray}
g(y)^{-1/4} a_1(x,y|A) &=&
 (\frac{1}{6}-\xi) R(x)
  -\frac{1}{2} (\frac{1}{6}-\xi)
 R_{;a}(x) z^{a}
- \frac{1}{3} {\cal A}_{ab}(x) z^{a}z^{b},	
\label{a1}\\
g(y)^{-1/4} a_2(x,y|A) &=& \frac{1}{2} (\frac{1}{6}-\xi)^2 R^2(x) 
-\frac{1}{3} {\cal A}_c^{c}(x)
\label{a2},
\end{eqnarray} 
where $z = y-x$  are Riemannian coordinates with the origin on the point $x$,
the semicolon indicates the covariant derivative and
\begin{eqnarray}
{\cal A}_{ab}(x) &:=& \frac{1}{2}(\frac{1}{6}-\xi) R_{;ab}(x)
 +\frac{1}{120}R_{;ab}(x) -\frac{1}{40}R_{ab;c}^{\:\:\:\:\:\:\:\:\:c}(x)
+\frac{1}{30} R_a^{\:\:\:c}(x) R_{cb}(x) \nonumber\\
& &-\frac{1}{60}R^{c\:\:\:d}_{\:\:\:a\:\:\:b}(x) 
R_{cd}(x) -\frac{1}{60} R^{cde}_{\:\:\:\:\:\:a}(x) R_{cdeb}(x) 
\label{abc}.
\end{eqnarray}

Let us consider the conformally coupled case, i.e. $\xi= 1/6$. Then
\begin{eqnarray}
 a_{1}(x|A) &=& 0,\\
a_{2}(x|A) &=& - \frac{1}{3} {\cal A}_c^c(x),\\
a_{1(a,b)}(x|A) &=& \frac{2}{3}{\cal A}_{ab}(x).
\end{eqnarray}
Employing (\ref{Gexplicit})  as well as the  coefficients above, we find
\begin{eqnarray}
3 (4\pi)^2 \: G_{ab}(x|A) =  2{\cal A}_{ab}(x) -  \frac{g_{ab}(x)}{2}\:
{\cal A}_c^c(x) \label{Gconformal}.
\end{eqnarray} 
It is obvious that, just like we expected 
\begin{eqnarray}
g^{ab}(x) \:G_{ab}(x|A) = 0\nonumber. 
\end{eqnarray}
As we said previously, this is related to the conformal anomaly. \\
Let us explicit the form of $G_{ab}(x|A)$. A few  trivial 
calculations\footnote{Taking also account
of the ``topological'' identity \cite{bd,fulling}\\
$\frac{1}{2}g_{ab}(x) R_{cdef}(x)R^{cdef}(x)
- 2 R_{acde}(x) R_{b}^{\:\:\:cde}(x) -4 \Delta R_{ab}(x) + 2 R_{;ab}(x)
+ 4 R_{ac}(x) R^{c}_{\:\:\:b}(x) 
-  4 R^{cd}(x) R_{cadb}(x) = - ^{(1)}H_{ab}(x) +4 ^{(2)}H_{ab}(x)$.}
produces the result
\begin{eqnarray}
G_{ab}(x|A) = \frac{1}{60 (4\pi)^2} \left[
{^{(2)}H_{ab}(x)}
-\frac{1}{3} {^{(1)}H_{ab}(x)}\right] \label{fine}.
\end{eqnarray}
The tensors  $^{(1)}H_{ab}(x)$ and $^{(2)}H_{ab}(x)$ are well-known conserved
tensors obtained by varying geometrical actions built up by quadratic 
curvature tensor terms. The right hand side of (\ref{fine}) is, up to 
constant overall factors, the
only linear combination of those tensor which is traceless.
Explicitly
\begin{eqnarray}
^{(1)}H_{ab}(x) &=& -\frac{1}{g^{1/2}} \frac{\delta}{\delta g^{ab}}\:
\int d^4x \: \sqrt{g} \: R^2(x) \nonumber\\
&=& 2 R_{;ab}(x) -2g_{ab}(x) \Delta R(x) + \frac{1}{2} g_{ab}(x)
R^2(x) - 2 R(x) R_{ab}(x), \nonumber
\end{eqnarray}
and
\begin{eqnarray}
^{(2)}H_{ab}(x) &=& -\frac{1}{g^{1/2}} \frac{\delta}{\delta g^{ab}}\:
\int d^4x \: \sqrt{g} \: R^{cd}(x)R_{cd}(x) \nonumber\\
&=& R_{;ab}(x) -\frac{1}{2} g_{ab}(x) \Delta R(x) -\Delta R_{ab}(x) 
 + \frac{1}{2} g_{ab}(x)
R^{cd}(x)R_{cd}(x) \nonumber \\
& & - 2 R^{cd}(x) R_{cdab}(x) \nonumber
\end{eqnarray}

We remind the reader that the term $\:\ln (\mu^2) \: G_{ab}(x|A)\:$ represents 
the  
finite renormalization part of our $\zeta$-function renormalization procedure.
The expression of the finite renormalization part we have found in (\ref{fine})
is exactly the same which appears in other regularization
and renormalization procedure (e.g. dimensional regularization) \cite{bd}.

\section{The simplest case: a box in the flat space}

Let us consider the simplest example of a super $\zeta$-regular theory.
That is a massless boson gas at the inverse temperature $\beta$ in a flat 
 box with a very large spatial volume $V$. This is 
the same example considered by Hawking in \cite{hawking} as far as
the $\zeta$-function regularization of the 
effective action was concerned; rather, we will deal with the stress
tensor. \\
For sake of simplicity, we  shall deal with the component $T_{00}$ of the 
stress tensor only.\\

The Euclidean action of the field is simply
\begin{eqnarray}
S = -\frac{1}{2} \:\int d^4x\:
 \delta^{ab}\partial_a \phi \partial_b \phi, \nonumber
\end{eqnarray} 
where $\delta^{ab}$ is the usually flat Euclidean metric. Notice that all 
coordinates define Killing vectors.
The stress tensor reads simply
\begin{eqnarray}
T_{ab}(x) = \partial_a\phi(x) \partial_b\phi(x)
 - \frac{1}{2}\delta_{ab}
\partial^c\phi(x) \partial_c\phi(x). \nonumber
\end{eqnarray}
We shall consider the box as a torus in order to use our method. 
 The motion operator is the 
trivial Laplacian with the sign changed $A = -\Delta$,
and we have the set of normalized eigenvectors
\begin{eqnarray}
\phi_{\vec{k},n}(x)  := \frac{e^{i\vec{x}\cdot \vec{k}
- ik_n x^0}}{\sqrt{\beta\: V}} 
, \label{seigenvectors}
\end{eqnarray}
where $\vec{x} \equiv
 (x^1,x^2,x^3)$ and $\vec{k} \equiv (k^1,k^2,k^3)$,
each $k^{i}$ being quantized by the usual torus quantization.  Also $k_n$ is
 quantized trivially by $k_n = 2\pi n/ \beta $ where $n = 0,\pm 1, \pm 2,...$.
Obviously, we have also
\begin{eqnarray}
A \phi_{\vec{k},n} = \lambda_{n,\vec{k}}\:
 \phi_{\vec{k},n}(x),\label{sequation}
\end{eqnarray}
where
\begin{eqnarray}
\lambda_{n,\vec{k}} := \vec{k}^2 + k_n^2.   \label{seigenvalues}
\end{eqnarray} 
The local zeta function reads
\begin{eqnarray}
\zeta(s,x|A) = (\beta V)^{-1}\: \sum_{n,\vec{k}} 
\: \lambda^{-s}_{n,\vec{k}} \label{slocal}
\end{eqnarray}
and finally, the $\bar{\zeta}_{00}(s,x|A)$ 
 function (see (\ref{barzabend}))
reads similarly
\begin{eqnarray}
\bar{\zeta}_{00}(s,x|A) = (\beta V)^{-1}\: \sum_{n,\vec{k}} 
\: 4\pi^2 \beta^{-2}
 n^2 \lambda^{-s}_{n,\vec{k}} \label{sz00}.
\end{eqnarray}
Proceeding as discussed in \cite{hawking}, we can rewrite the formulas above,
in the limit of a very large $V$
\begin{eqnarray}
\zeta(s,x|A) = \frac{4\pi}{(2\pi)^3\beta}\: 
\left\{ \int_\epsilon^{+\infty} dk k^{2-2s} \:+
2 \sum_{n=1}^{+\infty} \: \int_\epsilon^{+\infty} dk k^2 \: (4\pi^2 
\beta^{-2} n^2 + k^2)^{-s}\right\} \nonumber
\end{eqnarray}
and
\begin{eqnarray}
\bar{\zeta}_{00}(s,x|A) =
 \frac{16 \pi^4}{(2\pi)^3 \beta^3}\: 
\left\{ \int_\epsilon^{+\infty} dk k^{2-2s} \:+
2 \sum_{n=1}^{+\infty} \:n^2 \int_\epsilon^{+\infty} dk k^2 \: (4\pi^2 
\beta^{-2} n^2 + k^2)^{-s}\right\} \nonumber
\end{eqnarray}
The final results are (see \cite{hawking})
\begin{eqnarray}
\zeta(s,x|A) = -\frac{8\pi}{(2\pi)^3\beta}
(2\pi\beta^{-1})^{3-2s}\: \zeta_R(2s-3)\: (2-2s)^{-1}\: \frac{1}{2}
\:\frac{\Gamma(1/2)\Gamma(s-3/2)}{\Gamma(s-1)}, \label{szeta}
\end{eqnarray}
and (through (\ref{zabend}))
\begin{eqnarray}
\zeta_{00}(s+1,x|A) = \bar{\zeta}_{00}(s+1,x|A) -\frac{1}{2} \zeta(s,x|A)=
\nonumber 
\end{eqnarray}
\begin{eqnarray}
& & - \frac{32\pi^4}{(2\pi)^3\beta^3}
(2\pi\beta^{-1})^{1-2s}\: \zeta_R(2s-3)\: (-2s)^{-1}\: \frac{1}{2}
\:\frac{\Gamma(1/2)\Gamma(s-1/2)}{\Gamma(s)} \nonumber\\
& &+\frac{4\pi}{(2\pi)^3 \beta}
(2\pi\beta^{-1})^{3-2s}\: \zeta_R(2s-3)\: (2-2s)^{-1}\: \frac{1}{2}
\:\frac{\Gamma(1/2)\Gamma(s-3/2)}{\Gamma(s-1)}
\label{szetaab}.
\end{eqnarray}
We have dropped  parts dependent on the infrared cutoff $\epsilon$ by
putting
$\epsilon \rightarrow 0^+$ after one has fixed Re $s$ large finite, executed
the integration and performed the analytic continuation of this result
to $s=0$ (see \cite{hawking}).
$\zeta_R(s)$ is the usual Riemann zeta function which can be analytically
continued in the whole complex plane except for the only
singular point at $s=1$ .\\ 
We can analytically continue the functions above in the $s$-complex plane.  
In particular, notice that both functions can be analytically
continued in a neighborhood of $s=0$ {\em including} this point.  
 The apparent
 pole of $\zeta_{00}(s+1,x|A)$
 at $s=0$  is canceled out by the pole of $\Gamma(s)$ in the 
denominator; this means that $\zeta_{00}(s+1,x|A)$  takes no
poles in $s=0$ and defines a {\em super} $\zeta$-regular theory.
 Conversely, the $\zeta$ function  in (\ref{szeta}) vanishes at $s=0$.\\ 
As a final comment, we notice that the
 parameter $\mu$ will disappear from the  final  
renormalized effective action and the final renormalized $00$ component
of the stress tensor.\\
The $00$ component of the renormalized stress tensor can be now computed by
(\ref{regular}), taking the value at $s=0$ of the function in (\ref{szetaab}). 
We have
\begin{eqnarray}
-<T_{00}(x)>= -<T_0^0 (x)> = -<T_{L0}^{\:\:0} (x)> = 
-\zeta_{00}(1,x|A) = \frac{\pi^2}{30\beta^2}. \label{finet}
\end{eqnarray}
This is the well-known energy density of massless scalar bosons in a large 
box.\\
The well-known 
partition function can be  computed by the usual method through
$\zeta(s,x|A)$  and reads
 \cite{hawking} 
\begin{eqnarray}
Z_\beta = e^{\beta^{-3} \pi^2 V/90}. \label{ZH}
\end{eqnarray}
It is very simple to verify (\ref{thermal}) by using (\ref{finet}) and 
(\ref{ZH}).

\section{EINSTEIN'S CLOSED STATIC UNIVERSE}

The ultrastatic metric of the (Euclidean) 
Einstein closed static universe is \cite{bd}
\begin{eqnarray}
ds^2_{ECS} = d\theta^2 + g_{ij}dx^idx^j = d\theta^2 + a^2 \left( 
dX^2 + \sin^2X d\Omega^2_2 \right). \nonumber
\end{eqnarray}
$X$ ranges from $0$ to $\pi$ and $d\Omega^2_2$ is the usual metric
on $S_2$. The time coordinate $\theta$ ranges from $0$ to $\beta \leq +\infty$.
$\beta$ is the inverse temperature of the considered thermal state
referred to the Killing vector generated by the Lorentzian time
$i\theta$. The related
vacuum state corresponds to the  choice $\beta = +\infty$.
The curvature of the space is $R= 6/a^2$ and the Ricci tensor reads $R_{ij} =
2g_{ij}/a^2$, the remaining components vanish.\\
This manifold is closed, namely compact without boundary. Also
the spatial section at $\theta=$ constant are closed and their  volume  is
$V= 2\pi^2 a^3$. \\
Let us consider a conformally coupled massless scalar
field propagating within this manifold. We want to compute its stress tensor
referred to the thermal states  pointed out above, in particular we want to
get the vacuum stress tensor which is known in literature
\cite{bd}. Notice that all the required hypotheses
 to implement the stress-tensor $\zeta$-function approach
are  fulfilled: the Euclidean manifold is static and closed.\\
Let us build up the function $\zeta_{ab}(s,x|A)$ necessary to get
$\langle T_{ab}(x)\rangle_\beta$ through (\ref{general}) or (\ref{regular}).
The general expression of 
$\zeta_{ab}(s,x|A)$ is given in (\ref{zabend}). We can rewrite it down
 as
\begin{eqnarray}
\zeta_{ab}(s,x|A) &=& \bar{\zeta}_{ab}(s,x|A) 
-\xi\nabla_a\nabla_b \zeta(s,x|A) + \left(\xi - \frac{1}{4}
\right)g_{ab}(x) \Delta \zeta(s,x|A) \nonumber\\
& & + \xi R_{ab}(x) \zeta(s,x|A)
-\frac{1}{2} g_{ab}(x) \zeta(s-1,x|A), \label{zzbar}
\end{eqnarray}
where, in the sense of the analytic continuation of both sides in the
whole $s$ complex plane:
\begin{eqnarray}
\bar{\zeta}_{ab}(s,x|A) = {\sum_k}^{'} \lambda_k^{-s} \: \nabla_a\phi^*_k(x)
\nabla_b\phi_k(x) \label{zbarab}.
\end{eqnarray}
We are interested in the case $\xi=\xi_c:=1/6$ (conformal coupling 
in four dimensions).
The local $\zeta$ function is similarly given by
\begin{eqnarray}
\zeta(s,x|A) = {\sum_k}^{'} \lambda_k^{-s} \: \phi^*_k(x)
\phi_k(x) \label{zeta}.
\end{eqnarray}
The functions $\phi_k(x)$ define a normalized complete set of
eigenvectors of the Euclidean motion operator:
\begin{eqnarray}
A \phi_k = \lambda_k \phi_k\nonumber,
\end{eqnarray}
where, in our case
\begin{eqnarray}
A = -\partial^2_\theta -a^{-2} \Delta_{S_3} + \xi_c R\nonumber,
\end{eqnarray}
The explicit form of the considered eigenvalues and Kroneker's 
delta-normalized eigenvectors is
well-known \cite{bd}. In particular we have $k \equiv (n, q, l, m)$ where
$n = 0, \pm 1, \pm 2, \pm 3,...$, $q = 1, 2, 3,...$, $l = 0, 1, 2,..., q-1$,
 $m = 0, \pm 1,\pm 2,..., \pm l $ and 
\begin{eqnarray}
\lambda_k = \left(\frac{2\pi n}{\beta}\right)^2 + \left(\frac{q}{a}\right)^2
\label{eigenvalues}.
\end{eqnarray}
The following relations, which hold true for normalized eigenvectors, 
 are also useful. We leave the proofs of these to the reader.
\begin{eqnarray}
\sum_{lm} \phi_k^*(x)\phi_k(x) = \frac{q^2}{V\beta}, \label{firsto}
\end{eqnarray}
notice that the right hand side of the equation above 
is nothing but the  degeneracy of each eigenspace times $1/2\beta V$ (or
$1/\beta V$ when $n=0$);
\begin{eqnarray}
\sum_{lm} \partial_i\phi_k^*(x)\partial_j\phi_k(x) = g_{ij}(x)
\frac{q^2(q^2-1)}{3
V\beta a^2}, \label{secondA}
\end{eqnarray}
and ($x^0 := \theta$)
\begin{eqnarray}
\sum_{lm} \partial_0
\phi_k^*(x)\partial_0\phi_k(x) = \frac{(2\pi n q)^2}{V\beta^3} \label{third}.
\end{eqnarray}
We have also, because of the homogeneity of the space
\begin{eqnarray}
\zeta(s,x|A) =\frac{ \zeta(s|A)}{V\beta}, \label{local}
\end{eqnarray}
where $\zeta(s|A)$ is the global $\zeta$ function obtained
by summing over $\lambda^{-s}_k$ as usually
\begin{eqnarray}
\zeta(s|A) = {\sum_k}^{'} \lambda_k^{-s} \label{zetag}.
\end{eqnarray}
It is possible to relate the function $\bar{\zeta}_{ab}(s,x|A)$ to the function
$\zeta(s,x|A)$. Indeed, we notice that
\begin{eqnarray}
\lambda_k^{-s} \left(\frac{2\pi n}{\beta}\right)^2 =
\frac{\beta}{2(s-1)} \frac{\partial \lambda_k^{-(s-1)}}{\partial \beta}.
\nonumber 
\end{eqnarray}
The identity above inserted into the definition (\ref{zbarab}) for $a=b=0$,
taking (\ref{third}) into account, 
 yields
\begin{eqnarray}
\bar{\zeta}_{00}(s+1,x|A) = \frac{1}{2Vs} \frac{\partial\:\:\:}{\partial\beta}
\zeta(s|A), \label{zbar00}
\end{eqnarray}
or equivalently
\begin{eqnarray}
\zeta_{00}(s+1,x|A) = 
\bar{\zeta}_{00}(s+1,x|A)
 = -\frac{a}{2V\beta s} \frac{\partial\:\:\:}{\partial a}
\zeta(s|A) + \frac{\zeta(s|A)}{V\beta}, \label{zbar00'}
\end{eqnarray}
which follows from the  identity above taking account of
\begin{eqnarray}
2s\zeta(s|A) = \beta \frac{\partial\:\:\:}{\partial \beta}
\zeta(s|A) + 
a \frac{\partial\:\:\:}{\partial a}
\zeta(s|A). \label{symmetry}
\end{eqnarray}
The last identity is a simple consequence of the expression of the
eigenvalues (\ref{eigenvalues}).\\
Concerning the components $ij$ (the remaining components vanish)
we can take advantage from the identity
\begin{eqnarray}
\lambda_k^{-s} q^2 = 
\frac{3 a^3}{2(s-1)} \frac{\partial \lambda^{-(s-1)}_k}{\partial a} \label{id}.
\end{eqnarray}
Inserting this  into (\ref{zbarab}) for $a=i, b=j$, taking 
(\ref{secondA}) into account, it arises
\begin{eqnarray}
\bar{\zeta}_{ij}(s+1,x|A) = \frac{g_{ij}(x)}{3V\beta a^2}
 \left[ 
-\zeta(s+1|A) + \frac{a^3}{2s} \frac{\partial\:\:\:}{
\partial a}\zeta(s|A)\right]. \label{zbarij}
\end{eqnarray}
To get the renormalized stress tensor,
we have to compute $\zeta(s|A)$ or equivalently $\zeta(s,x|A)$ only.
The expansion  of the latter 
over the eigenvalues reads
\begin{eqnarray}
\zeta(s,x|A) &=& 
\frac{2}{V\beta}
\sum_{q=1}^{+\infty}
\sum_{n=1}^{+\infty}
q^2\left[ \left(\frac{2\pi n}{\beta}\right)^2+ 
\left(\frac{q}{a}\right)^2\right]^{-s} 
+ \frac{1}{V\beta}\sum_{q=1}^{+\infty}
q^2 \left[ \left(\frac{q}{a}\right)^2  \right]^{-s} \nonumber\\
&=&
\frac{2}{V\beta}
\sum_{q=1}^{+\infty}
\sum_{n=1}^{+\infty}
q^2\left[ \left(\frac{2\pi n}{\beta}\right)^2+ 
\left(\frac{q}{a}\right)^2\right]^{-s} 
+ \frac{a^{2s}}{V\beta} \zeta_R(2s-2) \label{z1}.
\end{eqnarray}
The last $\zeta$ function is  Riemann's one.\\
Let us introduce the Epstein function \cite{zerbinil}
 obtained by continuing (into a meromorphic function) 
the series in the variable $s$
\begin{eqnarray}
E(s,x,y) := \sum_{n,m=1}^{+\infty} \left( x^2 n^2 + y^2 m^2 \right)^{-s}
\label{epstein}.
\end{eqnarray} 
We get trivially
\begin{eqnarray}
\sum_{n,m=1}^{+\infty} m^2\left( x^2 n^2 + y^2 m^2 \right)^{-s}
= -\frac{1}{2y(s-1)}\frac{\partial\:\:\:}{\partial y}
E(s-1,x,y).
 \nonumber
\end{eqnarray}
Employing such an identity, we can rewrite the expression (\ref{z1}) of
$\zeta(s,x|A)$ as
\begin{eqnarray}
\zeta(s,x|A)
= \frac{a^{2s}}{V\beta} \zeta_R(2s-2) + \frac{a^3}{V\beta (s-1)}
\frac{\partial\:\:\:}{\partial a}
E(s-1,\frac{2\pi}{\beta},\frac{1}{a}). \label{z1medio}
\end{eqnarray}
no  expression of the Epstein function in terms of elementary functions 
 exists in
literature. Anymore, there exist a well-know expansion in terms of
MacDonald functions \cite{zerbinil}
\begin{eqnarray}
E(s,x,y)
&=& -\frac{1}{2} y^{-2s}\zeta_R(2s)
+ \frac{\sqrt{\pi} \Gamma(s-1/2)}{2x\Gamma(s)}
y^{1-2s} \zeta_R(2s-1) \nonumber\\
& &
+\frac{2\sqrt{\pi} x^{-2s}}{\Gamma (s)}
\sum_{m,n=1}^{+\infty} \left(\frac{\pi x m}{y n}
\right)^{s-1/2} K_{s-1/2}\left( \frac{2\pi y n m}{x}\right) \label{expansion1}.
\end{eqnarray}
Notice that, due to the negative exponential behavior 
of MacDonalds functions
$K_a(x)$ at large arguments, the last series defines a function 
which is analytic on the whole $s$ complex plane. The structure of the poles
of the Epstein function is due to the gamma and (Riemann's) zeta functions
in the first line of the formula above.
In particular there are only two simple poles at $s=1/2$ and $s=1$.\\
Taking account of the expression above and (\ref{z1medio}), we find
\begin{eqnarray}
\zeta(s,x|A) = \frac{\sqrt{\pi}}{4\pi V} \frac{\Gamma(s-3/2)}{\Gamma(s)}
(2s-3) a^{2s-1} \zeta_R(2s-3) 
-\frac{a}{V \Gamma(s)} \left(
\frac{\beta}{2\pi}\right)^{2s-2} \Xi(s, \beta/a), \label{z1fine} 
\end{eqnarray}
where the function $\Xi(s,\beta/a)$
given by
\begin{eqnarray}
\Xi(s,z)
= 2\pi \frac{d\:\:\:}{d z}
  \sum_{m,n=1}^{+\infty} \left(\frac{2\pi^2  m}{z n}
\right)^{s-3/2} K_{s-3/2}\left(n m z \right) \label{expansion},
\end{eqnarray}
 is analytic throughout the s complex
plane and, due to the 
large argument behavior
of the MacDonald functions,
 vanishes as $\beta \rightarrow +\infty $ like $(\beta/a)^{5/2-s}
\exp{-\beta/a}$ when Re $s \geq 0$.\\
Reminding the relation 
\begin{eqnarray}
2\frac{d\:\:\:}{d u} K_a(u)
= K_{a-1}(u) +K_{a+1}(u) 
\end{eqnarray}
the function $\Xi(s,z)$ and its $z$ derivative (see below) can be
evaluated numerically at the physical values $s=0$ and $s=1$
(see below).\\
The expression (\ref{z1fine}) which 
is very useful as far as  low temperature thermodynamics
in our manifold is concerned. Notice that, changing the role of 
$x$ an $y$ in the expression (\ref{expansion}), one may get an expression
for  $\zeta(s,x|A)$ useful at large temperatures.\\
Some remarks on (\ref{z1fine}) are in order. First notice that, 
due to the gamma functions into the denominators, $\zeta(s,x|A) \rightarrow 
 0$ like $s$ when $s\rightarrow 0$
and thus no trace anomaly appears and neither renormalization scale $\mu$
remains in the renormalized effective action. The found $\zeta$ function
is analytic throughout the $s$ complex plane except for the point $s=2$
where a simple pole appears.
Employing (\ref{zbar00}) and (\ref{zbarij}) we find that $\zeta_{ab}(s,x|A)$
is analytic at $s=1$ and thus the theory is a {\em super} $\zeta$-regular
theory.\\

Employing the definition (\ref{regular}), (\ref{zzbar})
 and the obtained expression for
$\zeta_{ab}(s,x|A)$, a few calculations lead us to
\begin{eqnarray}
\langle T_{La}^{\:\:b}(x)
 \rangle_\beta = \langle T_{a}^b(x) \rangle_\beta \equiv 
T(\beta) \:
(-1, \frac{1}{3},\frac{1}{3},\frac{1}{3}) \label{finalclosed},
\end{eqnarray}
where
\begin{eqnarray}
T(\beta) =
-\frac{1}{2V} \frac{\partial\:\:\:}{\partial\beta} \frac{\zeta(s|A)}{s}|_{s=0}
&=& \frac{1}{480 a^4 \pi^2} 
+ \frac{1}{a^4} \frac{d\:\:\:}{dz}|_{z=\beta/a} \frac{\Xi(0, z)}{z}
\label{T}.
\end{eqnarray}
Notice that the last  derivative term vanishes very fast
at low temperatures. \\
Now, one can  prove very simply  that the
 obtained stress tensor is conserved, has a vanishing trace and reduces to the
well-known  vacuum stress tensor in the closed Einstein universe \cite{bd}
as $\beta \rightarrow +\infty$
\begin{eqnarray}
\langle T_a^b(x) \rangle_{\scriptsize \mbox{vacuum}}
 \equiv \frac{1}{480 a^4 \pi^2} \:
(-1, \frac{1}{3},\frac{1}{3},\frac{1}{3}) \label{finalclosedvacuum}.
\end{eqnarray}
Taking account of $\zeta(0|A)=0$, we can rewrite (\ref{T}) as
\begin{eqnarray}
T(\beta) =
-\frac{1}{2V} \frac{\partial\:\:\:}{\partial\beta} \zeta'(0|A)
= -\frac{1}{V} \ln Z_\beta \nonumber
\end{eqnarray} 
where the prime means the $s$ derivative.
Hence, the relation (\ref{thermal}) holds true trivially.
The general relation between the Hamiltonian density and the stress-tensor
energy density in case of  static coordinates reads\footnote{Notice
that we are writing
 Lorentzian relations employing the Euclidean metric.
We could pass to use the  more usual Lorentzian 
metric simply through the identities
$g=-g_L$, $g_{00}= -g_{L00}$ and $g^{ij}= g^{ij}_L$.} 
\begin{eqnarray}
{\cal H} =   -T^0_0  + \xi g^{-1/2} \partial_i [g^{1/2}
(g^{ij} \partial_j \phi^2   \:- \phi^2 w^{i}) ] \label{TH},
\end{eqnarray}
  where $w^{a} = \frac{1}{2}
\nabla^{a} \ln g_{00}$. $w^{a}$ vanishes in the present case.
Let us employ such a relationship to evaluate the averaged value of the 
quantum Hamiltonian. We have to interpret (\ref{TH}) as
\begin{eqnarray}
\langle {\cal H} \rangle_\beta =   -
\langle T^0_0 \rangle_\beta + \xi g^{-1/2} \partial_i [g^{1/2}
(g^{ij} \partial_j \langle \phi^2 \rangle_\beta  \:- 
\langle \phi^2\rangle_\beta w^{i}) ] \label{TH'}.
\end{eqnarray}
As is well-known, provided the local $\zeta$ function is  regular
at $s=1$,  we can define
$\langle \phi^2(x) \rangle = \zeta(1,x|A)$. This is the case and we find
\begin{eqnarray}
\langle \phi^2(x) \rangle_\beta
 = -\frac{1}{48 \pi^2 a^2} - \frac{1}{2\pi^2 a^2}
\Xi(1,\beta/a). \nonumber
\end{eqnarray} 
This reduces to the known value as $\beta \rightarrow +\infty$ \cite{bd}. 
Notice that, due to the homogeneity of the space, there is not dependence on
$x$ and thus all derivatives in (\ref{TH'}) vanish yielding
$\langle {\cal H}\rangle_\beta =   -\langle T^0_0\rangle_\beta$.
Then (\ref{thermal}) can be  rewritten in terms of the averaged
Hamiltonian in the right hand side
\begin{eqnarray}
-\frac{\partial \ln Z_\beta
}{\partial\beta} \:
=  \langle H \rangle_\beta
 \label{thermalH}
\end{eqnarray}

\section{EINSTEIN'S OPEN STATIC UNIVERSE}

The ultrastatic metric of the (Euclidean)
Einstein closed static universe is \cite{bd}
\begin{eqnarray}
ds^2_{EOS} = d\theta^2 + g_{ij}dx^idx^j = d\theta^2 + a^2 \left(
dX^2 + \sinh^2X d\Omega^2_2 \right). \nonumber
\end{eqnarray}
$X$ ranges from $0$ to $+\infty$ and $d\Omega^2_2$ is the usual metric
on $S_2$. The time coordinate $\theta$ ranges from $0$ to $\beta \leq +\infty$.
Again, $\beta$ is the inverse temperature of the considered thermal state
referred to the Killing vector generated by the Lorentzian time
$i\theta$ and the  related
vacuum state corresponds to the  choice $\beta = +\infty$.
The curvature of the space is $R= -6/a^2$ and the Ricci tensor reads $R_{ij} =
-2g_{ij}/a^2$, the remaining components vanish.\\
This manifold is not closed and the spatial sections 
have not a finite volume.\\

Let us consider a conformally coupled massless scalar
field propagating within this manifold. As in the 
previously considered case, we want to compute its stress tensor
referred to the thermal states, in particular we want to
get the vacuum stress tensor. Notice that not all the required 
hypotheses  to implement the stress-tensor $\zeta$-function approach
are  fulfilled.   The manifold has no boundary but it is not compact.
We expect to find a continuous spectrum as far as the Euclidean 
motion operator is concerned.\\
However, we shall find that our method does work also in this case.
Notice that, now, we have to assume (\ref{general})
or (\ref{regular}) by definition and check on the obtained results
finally.

The form of the  eigenvalues $\lambda_k$ of the conformally coupled
massless Euclidean motion operator
\begin{eqnarray}
A = -\partial^2_\theta -a^{-2}\Delta_{H_3} + \xi_c R\nonumber,
\end{eqnarray}
 is
well-known \cite{bunch,bd}, 
we have, exactly as in the previous case 
\begin{eqnarray}
\lambda_k = \left(\frac{2\pi n}{\beta}\right)^2 + \left(\frac{q}{a}\right)^2
\label{eigenvaluesO},
\end{eqnarray}
where 
 $k \equiv (n, q, l, m)$ and
$n = 0, \pm 1, \pm 2, \pm 3,...$ , $q \in [0,+\infty)$,
$l= 0,1,2,3,...$
 $m = 0, \pm 1,\pm 2,..., \pm l $. The degeneracy depends only on
the indexes $l$ and $m$.\\
The following relations which hold true for  eigenvectors
$\phi_k(x)$ (which are Dirac's delta normalized in $q$ and Kroneker's delta
normalized in the remaining variables) 
 are also useful. We leave the proofs of these to the reader
(see also \cite{bunch}).
\begin{eqnarray}
\sum_{l,m} \phi_k^*(x)\phi_k(x) = \frac{q^2}{2\pi^2 a^3 \beta}, \label{firstO}
\end{eqnarray}
\begin{eqnarray}
\sum_{l,m} \partial_i\phi_k^*(x)\partial_j\phi_k(x) = g_{ij}(x)
\frac{q^2(q^2+1)}{6\pi^2 a^5 \beta }, \label{secondO}
\end{eqnarray}
and ($x^0 := \theta$)
\begin{eqnarray}
\sum_{l,m} \partial_0
\phi_k^*(x)\partial_0\phi_k(x) = \frac{(2\pi n q)^2}{2\pi^2 a^3\beta^3} 
\label{thirdO}.
\end{eqnarray}
Notice that the global $\zeta$ function simply does not exist because
the infinite spatial volume of the manifold.
Anyhow, we can compute the local $\zeta$ function as
\begin{eqnarray}
\zeta(s,x|A) := \int_0^{+\infty}dq\: \sum_{l,m,n} \phi_k^*(x)
\phi_k(x) \lambda_k^{-s} \label{localO}.
\end{eqnarray}
It is convenient to separate the contribution due to the terms with 
$n=0$ and introduce, as far as these terms are concerned, a cutoff
$\epsilon$ at low $q$. A few trivial manipulations of the expression
above yields
\begin{eqnarray}
\zeta(s,x|A) = \frac{a^{2s-3}}{2\pi^2 \beta}
\int_\epsilon^{+\infty} dq \: q^{2-2s}
+ \frac{1}{4\pi^2 \beta} \left(\frac{\beta}{2\pi}\right)^{2s-3} \zeta_R(2s-3)
\frac{\Gamma(1/2) \Gamma(s-3/2)}{\Gamma(s)}  \label{local2O}.
\end{eqnarray} 
The apparent divergent integral as $\epsilon \rightarrow 0^+$ can be made
harmless
as in \cite{hawking} putting 
$\epsilon \rightarrow 0^+$ after one has fixed Re $s$ large finite, executed
the integration and performed the analytic continuation of this result
to $s=0$.
This procedure generalize 
the finite volume prescription to drop the null eigenvalues in defining
the $\zeta$ function for the case  of an infinite spatial volume.
We have finally
\begin{eqnarray}
\zeta(s,x|A) = \frac{1}{8 \pi^2\sqrt{\pi} } \left( \frac{\beta}{2\pi}
\right)^{2s-4} \zeta_R(2s-3) \frac{\Gamma(s-3/2)}{\Gamma(s)} \label{localO3}.
\end{eqnarray}
Notice that $\zeta(0,x|A) = 0$ and thus no renormalization scale 
appears in the (infinite) partition function.\\
Let us evaluate $\bar{\zeta}_{ab}(s,x|A)$. The only nonvanishing 
components are $00$ and $ij$. In the first case we have directly from the 
 definitions
(omitting the terms with $n=0$ as above)
\begin{eqnarray}
\zeta_{00}(s+1,x|A) &= & \bar{\zeta}_{00}(s+1,x|A)
 =
\int dq \sum_{l,m,n} \left(\frac{2\pi n}{\beta}\right)^2
 \lambda_k^{-s} \phi_k^*(x)\phi_k(x)\nonumber\\
&=& \frac{1}{8\pi^2\sqrt{\pi}} \left( \frac{\beta}{2\pi}
\right)^{2s-4} \zeta_R(2s-4) \frac{\Gamma(s-1/2)}{\Gamma(s+1)}.
\label{zbar00O}
\end{eqnarray}
In order to compute the remaining components of $\bar{\zeta}_{ab}$ we can use 
(\ref{secondO}) and the relation in (\ref{id}) once again.
We find
\begin{eqnarray}
\bar{\zeta}_{ij}(s+1,x|A) = \frac{1}{3a^5} g_{ij}(x)
\zeta(s+1,x|A) + \frac{1}{2s} g_{ij}(x) \zeta(s,x|A). 
\end{eqnarray}
We have found  that $\zeta_{ab}(s,x|A)$ is analytic in $s=1$, hence
the theory is a {\em super} $\zeta$-regular theory once again. We can use 
(\ref{regular}) to compute the stress tensor.\\
Through (\ref{zzbar}) and (\ref{regular}) we find finally
\begin{eqnarray}
\langle T_{La}^{\:\:b} \rangle_\beta =\langle T_a^b \rangle_\beta 
\equiv
T(\beta) (-1,\frac{1}{3},\frac{1}{3},\frac{1}{3})  \label{tabO},
\end{eqnarray}
where
\begin{eqnarray}
T(\beta) = \frac{\pi^2}{30\beta^4} \label{tbetaO}.
\end{eqnarray}
The stress tensor in (\ref{tabO}) is conserved and  traceless
as we expected from the general theory. $\langle T_a^b \rangle_\beta$
vanishes as $\beta \rightarrow +\infty$, this agrees with the known
result \cite{bd} that the stress tensor in the vacuum state of the
open Einstein universe vanishes.\\
Notice that the found stress tensor,
in the considered 
components, is exactly the same than in Minkowski spacetime.\\

Let us finally consider (\ref{thermal}). In this case the left hand side of 
(\ref{thermal})
does not exist because that simply diverges.
Nevertheless, we can notice that the divergence of the partition function
is due to the volume divergence only and 
the remaining factor does not depend on the position on the spatial section,
namely
\begin{eqnarray}
\ln Z_\beta = V \ln {\cal Z}_\beta = V \times \left(\beta \frac{1}{2}
\zeta'(0,x|A)\right) = V \times \frac{\pi^3}{90\beta^3}  \label{xyz}
\end{eqnarray}
where $V$ diverges and, actually, $\zeta'(0,x|A)$ 
does not depend on $x$ due to the
homogeneity of the spatial manifold. 
This is the same situation than arises in the Minkowski spacetime.\\
 We expect that,
although (\ref{thermal}) does not make sense, a local version could yet
make sense. Indeed, one can get very simply from (\ref{tabO}) 
and (\ref{xyz}) 
\begin{eqnarray}
-\frac{\partial \:V \ln {\cal Z}_\beta}{\partial\beta} 
= - V \: \langle T_{0}^{0}\rangle_\beta =
 -\int_V d\vec{x} \:\sqrt{g}\: \langle T_{L\: 0}^{\:\:0}(\vec{x})
\rangle_\beta
\end{eqnarray}
on any finite volume $V$.
 As in the previously discussed case,
$\langle \phi^2(x) \rangle_\beta$ 
can be obtained  by evaluating the local $\zeta$ function
at $s=1$, we get
\begin{eqnarray}
\langle \phi^2(x) \rangle_\beta = \frac{1}{12\beta^2}.
\end{eqnarray}
Notice that this vanishes as $\beta \rightarrow +\infty$
namely, in the vacuum state as is known \cite{bunch}. Furthermore, it 
does not depend on $x$ and thus, through (\ref{TH}) and noticing that $w^{a}
=0$
(see the Einstein closed universe case), 
$\langle T_0^0 \rangle_\beta = -\langle {\cal H} \rangle_\beta $.
We can write finally, with an obvious meaning
\begin{eqnarray}
-\frac{\partial \:V \ln {\cal Z}_\beta
}{\partial\beta} \:
=  \langle H_V \rangle_\beta.
 \label{thermalH2}
\end{eqnarray}

\section{THE CONICAL MANIFOLD}

Let us consider the Euclidean manifold ${\cal C}_\beta \times {\cal R}^2$
 endowed with the metric
\begin{eqnarray}
ds^2 = r^2 d\theta^2 + dr^2 + dz_1^2 + dz_2^2 \label{conic}, 
\end{eqnarray}
where $(z_1,z_2)\in {\cal R}^2$, $r\in [0,+\infty)$, $\theta \in [0,\beta)$
when $0$ is identified with $\beta$. ${\cal C}_\beta \times {\cal R}^2$
is a cone with deficit angle given by $2\pi - \beta$.
That is the Euclidean manifold corresponding to the finite temperature
($T= 1/\beta$) quantum field theory in the Rindler space. In such a case
$\theta$ is the Euclidean time of the theory.
This is also a good
approximation of a large mass black hole near the event horizon.
Equivalently, considering $z_1$ as the Euclidean time, the metric above 
defines the Euclidean section (at zero temperature) of a cosmic string
 background. In this case $(2\pi-\beta)/8\pi G$ is the mass of the string.\\
The metric in (\ref{conic}), considered as the Rindler Euclidean metric,
 is static but not {\em ultrastatic}. Another important point is that
such a metric is not homogeneous in the spatial section.\\
The considered manifold is flat everywhere except for  
 conical singularities which appear at $r=0$ whenever $\beta \neq 2\pi$.
These singularities produces well-known 
Dirac's delta singularities in the curvatures
of the manifolds at $r=0$ \cite{singular}. The physics involved in such
anomalous curvature it is not completely known. Actually, we shall see 
shortly that
one can ignore completely the anomalous curvature dealing with the 
stress tensor renormalization also considering nonminimal coupling with
the scalar curvature.\\
As is well-known, the particular value $\beta_H = 2\pi$ defines
 the Hawking-Unruh temperature in the Rindler/large-mass-black-hole
interpretation, the corresponding thermal state being nothing but the
Minkowski vacuum/Hartle-Hawking (large mass) vacuum.\\
The thermal Rindler stress tensor (renormalized with respect to the 
Minkowski vacuum)
 which coincide, in the Euclidean approach, to
  the zero-temperature cosmic-string stress
tensor (renormalized with respect to the Minkowski vacuum) has been
computed by the point splitting approach \cite{frolov}.\\
Such results has been only partially reproduced by some
 $\zeta$-function or (local) heat kernel
approach \cite{kcv,ielmo}. This is because these approaches were employed to
renormalize the effective action only, and thus the stress tensor
was computed assuming further hypotheses on its form or assuming 
some statistical-mechanical law as holding true \cite{kcv,ielmo}.\\
 Recently, in \cite{iellici}, also
 the massive case has been considered by employing an off-diagonal 
$\zeta$-function approach and a subtraction procedure similar to
that is employed within  the point-splitting framework.\\
Here, we shall consider the massless case only. We shall check our approach
for every value of the curvature coupling proving that the same results
got by the point-splitting approach naturally arise.
The important point is that, due to the complete independence of the
method from statistical mechanics, we shall be able to discuss
 the statistical mechanics meaning (if it exists)
 of our results {\em a posteriori}.\\

Let us consider first the case of the {\em minimal} coupling $\xi=0$.
This avoids all problems involved dealing with the singular curvature
on the tip of the cone generated by the conical singularity.
The  
$\zeta$ function of the effective action  in  conic 
backgrounds 
 has been computed by several authors \cite{zcv} 
 also in the massive
scalar case \cite{iellici}
and for photons and gravitons \cite{moiel}. \\
Discarding the singular curvature by posing 
$\xi=0$, a
 complete normalized  set of  eigenvectors of the massless Euclidean motion
 operator\footnote{We are considering a particular 
self-adjoint extension of the formally
self-adjoint Laplace-Beltrami operator in the conical manifold.
The general theory of these extensions has been studied in \cite{kay}.}
 $A = -\Delta_{{\cal C}_\beta \times {\cal R}^2}$ 
 is \cite{zcv} 
\begin{eqnarray}
\phi_q(x) = \frac{1}{2\pi} \sqrt{\frac{\lambda}{\beta}}\:
e^{ikz} e^{i\frac{2\pi n}{\beta} \theta}
J_{\frac{2\pi |n|}{\beta}}(\lambda r) \label{eigenvector3}
\end{eqnarray}
where $z=(z_1,z_2)$,
 $q = (n,k,\lambda)$, $n= 0, \pm 1,\pm 2,...$, $k= (k_1,k_2)\in {\cal R}^2$
$\lambda \in [0, +\infty)$.
The considered eigenfunctions are Kroneker's delta normalized in the index
$n$ and Dirac's delta normalized in the remaining indices.
The corresponding eigenvalues are
\begin{eqnarray}
\lambda_q = \lambda^2 + k^2  \label{eigenvalues3}.
\end{eqnarray}
The $\zeta$ function of $A$ has been computed explicitly and reads
\begin{eqnarray}
\zeta(s,x| A) = \frac{r^{2s-4}}{4\pi \beta \Gamma(s)} I_\beta(s-1)
\label{zeta3}.
\end{eqnarray}
$I_\beta(s)$ is a well-known meromorphic function \cite{zcv}
carrying a simple pole at $s=1$. Known values are also
\begin{eqnarray}
I_\beta(0) &=& \frac{1}{6\nu}(\nu^2 - 1) \label{I0},\\
I_\beta(-1)&=& \frac{1}{90 \nu}(\nu^2 -1)(\nu^2 +11) \label{I-1},
\end{eqnarray} 
where we defined $\nu := 2\pi/\beta$.\\
Notice that $\zeta(0,x|A)=0$ and thus no scale remains into the renormalized
local effective action. $\langle \phi^2(x) \rangle$ can be computed 
by evaluating the local $\zeta$ function at $s=1$.\\
The function $\bar{\zeta}_{ab}(s,x|A)$
 can be computed making use of intermediate 
results contained in \cite{iellici}. A few calculations lead us to
\begin{eqnarray}
\bar{\zeta}_{\theta\theta}(s,x|A) &=&  
\frac{r^{2s-4}\Gamma(s-3/2)}{4\pi \sqrt{\pi} \beta \Gamma(s)} H_\beta(s-1)
\label{thetatheta3},\\
\bar{\zeta}_{rr}(s,x|A) &=&  
\frac{1}{2r}\partial_r r\partial_r \zeta(s,x|A)
 - \frac{1}{r^2}
\bar{\zeta}_{\theta\theta}(s,x|A) + 4\pi (s-2) \zeta_{D=6}(s,x|A),
 \label{rr3}\\
\bar{\zeta}_{z_1z_1}(s,x|A) &=&  \bar{\zeta}_{z_2z_2}(s,x|A) = 
2\pi \zeta_{D=6}(s,x|A). \label{zz3}
\end{eqnarray}
All remaining components vanish.
The meromorphic function $H_\beta(s)$ has been defined in \cite{iellici},
it has a simple pole at $s=2$ and known values are
\begin{eqnarray}
H_\beta(0) &=& \frac{1}{120\nu}(\nu^4 - 1) \label{H0},\\
H_\beta(1)&=& -\frac{1}{12 \nu}(\nu^2 -1) \label{H1}.
\end{eqnarray} 
The function $\zeta_{D=6}(s,x|A)$ is the $\zeta$ function of the 
effective action in ${\cal C}_\beta\times {\cal R}^4$ \cite{iellici}, it reads
\begin{eqnarray}
\zeta_{D=6}(s,x| A) = \frac{r^{2s-6}}{(4\pi)^2 \beta \Gamma(s)} I_\beta(s-2)
\label{zeta6}.
\end{eqnarray}
From the above equations and (\ref{zzbar}) it follows that
$\bar{\zeta}_{ab}(s,x|A)$ is analytic at $s=1$ and thus
 the theory is 
super $\zeta$-regular once again. Hence, we can use (\ref{regular}) to
compute the stress tensor.  Trivial calculations employing (\ref{zzbar})
with $\xi=0$ and (\ref{regular}) produce
\begin{eqnarray}
\langle T_a^b(x)_{\xi=0} \rangle_\beta  &\equiv& 
\frac{1}{1440\pi^2 r^4}
\left\{ \left[ \left(\frac{2\pi}{\beta} \right)^4 -1 
\right]\mbox{diag}(-3,1,1,1)\right.\nonumber\\
& &- \left.
20\left[ \left(\frac{2\pi}{\beta} \right)^2 -1 
\right]\mbox{diag}(\frac{3}{2},-\frac{1}{2},1,1)
\right\} \label{xizero}.
\end{eqnarray}
This is the correct result arising by the point-splitting approach
\cite{frolov} in the case of the minimal coupling. Let us prove that our method
reproduce also the remaining cases. \\
In general, the relationship  between  the minimally coupled stress-tensor
and the generally coupled stress tensor can be trivially obtained 
by varying the action containing the usual coupling with the curvature,
it  reads
\begin{eqnarray}
T_{ab}(x)_\xi = T_{ab}(x)_{\xi=0} + \xi 
\left[ (R_{ab}- \frac{1}{2}g_{ab}R) \phi^2(x) + g_{ab} \Delta \phi^2(x) 
- \nabla_a \nabla_b \phi^2(x) \right] \label{class}.
\end{eqnarray}
It is worthwhile stressing that the last $\xi$-parametrized term appears also
when the manifold is flat.
We can interpret quantistically  this relationship as
\begin{eqnarray}
\langle T_{ab}(x)_\xi\rangle  = \langle T_{ab}(x)_{\xi=0}\rangle + \xi
\langle Q(x)_{ab}\rangle, \nonumber
\end{eqnarray}
where
\begin{eqnarray} 
\langle Q(x)_{ab}\rangle
 := \left[ (R_{ab}(x)- \frac{1}{2}g_{ab}(x)R(x)) \langle \phi^2(x)
\rangle + g_{ab} \Delta \langle \phi^2(x) \rangle 
- \nabla_a \nabla_b \langle \phi^2(x)\rangle \right] \label{Q}.
\end{eqnarray}
Now 
$\langle T_{ab}(x)_{\xi=0}\rangle_\beta$
 is known by (\ref{xizero}), $R_{ab}(x)=0$, $R(x)=0$ and thus we can 
compute $\langle T_{ab}(x)_{\xi}\rangle_\beta$
employing the known value of 
$\langle \phi^2(x)\rangle_\beta$. We have,
through  (\ref{zeta3})
\begin{eqnarray}
\langle \phi^2(x)
\rangle_\beta = \zeta(1,x|A) = 
\frac{1}{48\pi^2 r^2}\left[\left(\frac{2\pi}{\beta}
\right)^2-1 \right] \label{phi3}.
\end{eqnarray}
The final result is exactly that of the point-splitting approach: 
\begin{eqnarray}
\langle T_{La}^{\:\:b}(x)_{\xi} \rangle_\beta  =
\langle T_a^b(x)_{\xi} \rangle_\beta  &\equiv& 
\frac{1}{1440\pi^2 r^4}
\left\{ \left[ \left(\frac{2\pi}{\beta} \right)^4 -1 
\right]\mbox{diag}(-3,1,1,1)\right.\nonumber\\
& &+ \left.
20(6\xi -1)\left[ \left(\frac{2\pi}{\beta} \right)^2 -1 
\right]\mbox{diag}(\frac{3}{2},-\frac{1}{2},1,1)
\right\} \label{xigen}.
\end{eqnarray}
The same result arises by employing the definition of
 $\zeta_{ab}(s,x|A)$ given in (\ref{zzbar}) with the chosen value of 
$\xi$, provided $\bar{\zeta}_{ab}(s,x|A)$
and $\zeta(s,x|A)$ are those computed in the {\em minimal coupling case}.
This means that, concerning the renormalization of the stress tensor,
the presence of the conical singularity which determines
a singular curvature on the tip of the cone  is completely irrelevant.
Concerning the quantum state, there is no difference between different
 couplings with the curvature. The 
$\xi$-parametrized  
term remains as a relic 
in the stress tensor because of the {\em classical} formula (\ref{class}).
This term does not come out
 from the quantum state once one
fixed the renormalization procedure.  
We see that the renormalization of the stress tensor can be managed completely
 by our Euclidean $\zeta$-function
 approach on the physical manifold instead of the {\em optical} manifold 
 not depending on the presence of the conical singularity
in the Euclidean manifold.\\
The knowledge of the averaged and renormalized stress tensor makes us able 
to compute the averaged and renormalized Hamiltonian of the system. 
 The Hamiltonian of the theory should not depend on the parameter $\xi$
because that cannot appear into the Lorentzian action the manifold
being flat.  Notice that 
there is no conical singularity in the Lorentzian
 theory! 
Not depending on $\xi$, 
the classical Hamiltonian density coincides with the changed sign  
energy component of the
stress tensor in the minimal coupling.
Indeed, employing (\ref{TH}), we can write down
\begin{eqnarray}
 \langle {\cal H}(x) \rangle_\beta   
= -\langle T^0_0(x)_{\xi=0}\rangle_\beta = 
\frac{3}{1440\pi^2 r^4}
\left[  \left(\frac{2\pi}{\beta} \right)^4 
 - 10 \left(\frac{2\pi}{\beta} \right)^2 -11  \right] \label{xizero1}.
\end{eqnarray}\\
Let us finally 
consider the problem of  the validity of the relation (\ref{thermal})
 in some sense.
The spatial section  is neither finite nor homogeneous, we could have 
problems with the use of cutoffs. It is not obvious that such a relation as
(\ref{thermal}) can hold true in our case considering cutoff smeared quantities
as\footnote{Notice that also the area $A$ of the horizon is a cutoff because
the actual area is infinite. This cutoff is a trivial overall factor.
We shall omit this cutoff as an index in the 
following formulae for sake of simplicity.}
\begin{eqnarray}
\ln Z_{\beta \epsilon} 
&:=& \int_{r>\epsilon} d^4x \sqrt{g}
\:\frac{1}{2}\frac{d\:\:\:}{ds}|_{s=0}\zeta(s,x|A), \label{zed}\\
{\cal Q}_{\epsilon}(\beta)
&:=&  \int_{r>\epsilon} d^3x \sqrt{g} 
\langle Q_0^0(x)\rangle_\beta,\\
\langle H_\epsilon \rangle_\beta &:=& 
\int_{r>\epsilon} d^3x \sqrt{g} 
\langle {\cal H}\rangle_\beta
\end{eqnarray}
and finally 
\begin{eqnarray}
{\cal E}_{\epsilon\xi}(\beta):=
  -\int_{r>\epsilon} d^3x \sqrt{g} 
\langle T_0^0(x)_\xi\rangle_\beta
 =  \int_{r>\epsilon} d^3x \sqrt{g} 
\langle {\cal H}\rangle_\beta
 - \xi {\cal Q}_{\epsilon}(\beta).
\end{eqnarray}
In particular we have from (\ref{zeta3})
\begin{eqnarray}
\ln Z_{\beta \epsilon} = \frac{A\beta}{2880 \pi^2 \epsilon^2}
\left[\left(\frac{2\pi}{\beta}\right)^4
 + 10 \left(\frac{2\pi}{\beta}\right)^2 -11 \right],
\label{abcdo}
\end{eqnarray}
where $A$ is the area of the event horizon, the regularized volume 
of the spatial section is $V_\epsilon = A/(2\epsilon^2)$.
Notice that, actually, the conserved charge ${\cal Q}_\epsilon(\beta)$
is a boundary integral which diverges on the conical singularity.
Indeed, it can be expressed by the integration of (\ref{TH}) and it should
 discarded if the manifold were regular.
Notice that the choice of values of  $\xi$ determines  different values
of ${\cal E}_{\xi\epsilon}$  due to the $\xi$-parametrized
boundary term $\xi {\cal Q}_\epsilon$
 in the stress tensor. Conversely, $\ln Z_{\beta\epsilon}$
does not depend on $\xi$.\\
 If something like (\ref{thermal})
holds true for a fixed value of $\epsilon$,
 it does just for a particular and unique value of $\xi$. Actually 
few calculation through  (\ref{xigen})
prove that, {\em not  depending on the value of} $\epsilon$
\begin{eqnarray}
- \frac{\partial \ln Z_{\beta \epsilon}}{\partial\beta} 
= {\cal E}_{\epsilon \xi=1/9}(\beta) + {\cal E}_{\epsilon} =
 \langle H_\epsilon \rangle_\beta - \frac{1}{9}{\cal Q}_\epsilon(\beta)
+ {\cal E}_\epsilon.
\end{eqnarray}
The last term is an opportune
 constant energy
\begin{eqnarray}
{\cal E}_{\epsilon} = \frac{A}{120 \pi^2 \epsilon^2}. \nonumber
\end{eqnarray}
The presence of such an added constant
could be expected from the fact that  the energy
${\cal E}_{\epsilon\xi}$ is renormalized to vanish at $\beta=2\pi$ instead of 
$\beta= +\infty$. Conversely, there 
is no trivial explanation of the presence of
the $\beta$-{\em dependent}  term $- \frac{1}{9}{\cal Q}_\epsilon(\beta)$.
Then, in the considered case,
in the right hand side of (\ref{thermal}) 
does not appear the Hamiltonian which, at least 
classically, corresponds to the value $\xi=0$ as discussed above.\\
One could wonder whether or not $Z_{\beta \epsilon}$
defined in  (\ref{zed}) can be considered a (regularized) partition function
of the system. The simplest answer is obviously not because a fundamental
relationship of statistical mechanics does not hold true.\\
In general, one could think that this negative result 
arises because we have dropped a contribution due to the conical singularity.
This singularity produces a Dirac delta in the curvature on the tip of the
cone in the Euclidean manifold. The integral of the Lagrangian get a 
contribution from this term in the case of a nonminimal coupling with the
curvature. 
The problem of the contributions of these possible terms, in particular
in relation to the black-hole entropy 
has been studied by several authors
(see \cite{FF,FFZ,BS,S,H,M} and references therein), anyhow,
in this paper we shall not explore such a possibility.\\
In any cases, it is worthwhile stressing that
 the found Euclidean effective action (\ref{abcdo})
is the correct one in order to get the {\em thermal}
 renormalized stress tensor by (formal)
variation with respect to the background metric.
We re-stress that the obtained stress tensor is exactly that obtained by the 
point-splitting approach.\\

The question whether or not the effective action computed by the $\zeta$ 
function defines also  the logarithm of the partition function 
(renormalized with respect to the Minkowski vacuum)
is not a simple question. \\
The problem is interesting on a physical ground 
also because the partition function of the field around a black hole
(we remind the reader that the Rindler metric represents a large mass
black hole) is used to compute the quantum corrections to the
Bekenstein-Hawking entropy as early suggested by 't Hooft
\cite{hooft}
or  to give a reason for the complete B-H entropy in the 
framework of the induced gravity considering massive fields
nonconformally coupled \cite{FF,FFZ,FF2}. \\
As noticed  in {\bf Section VI},
 on a more general ground, the considered  problem is also interesting
because
there exist two  not completely equivalent approaches
 to implement the statistical mechanics 
of a quantum field in a curved spacetime through the use of a path
integral techniques and, up to the knowledge of the author,
there is not a definitive choice of the method.
In this work, we have employed the path integral
in the physical manifold instead of in the {\em optical} related manifold.
We remind the reader that in the case of a {\em static}
 but not {\em ultrastatic} 
spacetime, the naive approach based on the phase-space path
integral leads one to a definition of the partition function as an Euclidean
path integral performed in the configuration space within the
{\em optical}
 manifold\footnote{This is the ultrastatic manifold
conformally related to the physical manifold
by defining the optical metric through
$\tilde{g}_{ab} := g_{ab}/g_{00}$. The Euclidean action employed on the
optical manifold is the physical action conformally transformed (including
the matter fields)
following the conformal transformation written above.} 
instead of the physical one \cite{dealwis}. Other approaches
\cite{toms} lead one to the definition of the partition function as a path 
integral in the physical manifold. \\
When the spatial section of the space is 
regular (e.g. closed) and thus the 
path integral regularized through  the $\zeta$-function approach
yields a finite result, formal manipulations of the path integral 
prove that   these two different definitions lead to the same result
up to the renormalization of the zero point energy  \cite{report}.
 In such a case 
 these definitions are substantially 
equivalent.
 When the manifold is not regular, e.g. it has spatial
sections with an infinite volume or has boundaries, in principle one may
loose   such an equivalence.
Indeed,
as far as the effective actions are concerned  in our case  we have 
\begin{eqnarray}
\ln Z_{\beta \epsilon}
 = \frac{A\beta}{2880 \pi^2 \epsilon^2}
\left[\left(\frac{2\pi}{\beta}\right)^4
 + 10 \left(\frac{2\pi}{\beta}\right)^2 -11 \right],
\nonumber
\end{eqnarray}
and 
\begin{eqnarray}
\ln Z^{\scriptsize \mbox{opt}}_{\beta \epsilon} =
 \frac{A\beta}{2880 \pi^2 \epsilon^2}
\left(\frac{2\pi}{\beta}\right)^4
\label{abcsopt}
\end{eqnarray}
The latter result can be directly obtained  noticing that the optical
manifold of the Rindler space is the open Einstein
static universe \cite{bd}. Hence the latter effective action above is nothing
but that computed previously
in the open Einstein universe (in the conformal coupling).\\
Considering the  effective action computed as a path integral in the optical
manifold we have
\begin{eqnarray}
- \frac{\partial 
\ln Z^{\scriptsize \mbox{opt}}_{\beta \epsilon}}{\partial\beta} 
= {\cal E}_{\epsilon\xi=1/6}(\beta) + {\cal E}_\epsilon'           
\label{xyzt}
\end{eqnarray}
 One could conclude that,  once again, 
there is not the Hamiltonian in 
the right hand side, also discarding the constant energy. Actually,
 this result involves more subtle considerations. Indeed, we shall prove that
this naive conclusion is not correct. \\
Let us suppose to implement  canonical QFT \cite{bd}
for a massless field conformally coupled
 directly on the optical manifold,
namely in the open Einstein static universe as it were the physical manifold. 
Obviously, we should
get exactly the effective action which appears in (\ref{xyzt}).
Furthermore,
  (\ref{xyzt}) is nothing but (\ref{thermalH}) and the right hand side of
(\ref{xyzt}) is nothing but the averaged $\epsilon$-regularized  
Hamiltonian of the QFT in the open Einstein universe.
 Such a Hamiltonian can be also
obtained as a thermal average of the Hamiltonian operator got from the
 canonical QFT employing the {\em normal order prescription}
and
employing the usual definition of the partition function (summing the
Boltzmanian exponential in the  energy levels of the
states in the canonical ensemble) \cite{mova}.\\ 
Implementing  the canonical quantization in the Rindler space for a
 massless scalar field, one trivially
finds that  an isomorphism exists between the Fock space built up on the
Fulling-Rindler vacuum and the Fock space built up on the natural vacuum of the
QFT in the open Einstein static universe (in the conformal coupling). Indeed, 
this isomorphism arises from the conformal relationship between
the wavefunctions of the particles related to the quantized 
fields. This relation defines a one-to-one map from the one-particle Hilbert
space of the Einstein open universe  to the one-particle Hilbert space
of the Rindler space which maintains the value 
of the corresponding indefinite  scalar products \cite{bd}.
 This map defines a unitary
isomorphism between the two Fock spaces provided one require that this
isomorphism transform the vacuum state of the Einstein open universe into
the Fulling-Rindler vacuum.
In particular, also the Hamiltonian operators are unitarily identified
provided one use the normal order prescription in both cases.\\
As a result we find that
 the right hand side of (\ref{xyzt}) coincides  also  with the averaged
Hamiltonian operator built up in the framework of the canonical quantization
in the Rindler space with respect to the Fulling-Rindler vacuum!\\
In this sense (\ref{xyzt}) is the usual statistical mechanical relationship
between the canonical energy and the partition function in the Rindler 
space.\\
The central point is that the renormalization scheme employed is the
normal order prescription with respect to the Fulling-Rindler vacuum and
not the point-splitting procedure.\\ 
We can finally 
compare the averaged Rindler Hamiltonian of the canonical quantization 
$\langle H_\epsilon^{\scriptsize \mbox{can}} 
\rangle_\beta$ which is renormalized by the {\em normal order prescription}
in the the Fulling-Rindler vacuum 
 with the averaged Rindler Hamiltonian $\langle H_\epsilon \rangle_\beta$
obtained by integrating (\ref{xizero1}).
The latter  is renormalized with respect the
 Minkowski 
vacuum  by the {\em point-splitting procedure}. 
 We find
\begin{eqnarray} 
\langle H_\epsilon \rangle_\beta - \langle 
H_\epsilon^{\scriptsize \mbox{can}}  \rangle_\beta =
-\frac{3}{2880\pi^2 \epsilon^2} -\frac{30}{2880\pi^2 \epsilon^2} 
\left[ \left(\frac{2\pi}{\beta} \right)^2 -1 
\right] = -\frac{1}{960 \pi^2 \epsilon^2} - \frac{1}{6}
 {\cal Q}_\epsilon(\beta). \label{eqfine}
\end{eqnarray}
The first term in the right hand side is trivial: it takes account of the
difference of the zero-point energy. The second term is quite unexpected.
It proves that the point-splitting procedure 
 (or equivalently our $\zeta$-function procedure)
to renormalize the stress
tensor and hence the Hamiltonian
is not so trivial as one could expect, this is 
 because it involves terms which do not represent a trivial
zero-point energy renormalization.\\
Concerning the conical manifold, the conclusion is that the theory 
in the optical manifold leads us naturally 
to an effective action which can be considered the logarithm
of the partition function provided we renormalize the theory with respect 
to the Fulling-Rindler vacuum.
 Conversely, the effective action evaluated in the 
physical manifold is the correct effective action which produces
the thermal stress tensor
by formal variations with respect to the metric.
 This stress tensor
is that obtained also by the point-splitting procedure and thus
renormalizing with respect to the Minkowski vacuum.

\section{SUMMARY}

In this paper we have presented a new approach to renormalize the one-loop
stress tensor in a curved background based on an opportune $\zeta$-function 
regularization. The procedure has been developed within the Euclidean 
formalism and in the hypothesis of a closed manifold and a real
scalar field.\\
We do not think that our approach should change dramatically relaxing
such hypotheses. This is because the same $\zeta$-function approach 
to renormalize the effective action  was born in a similar context and 
has been successively developed into a very general context.
In fact, we have used the method also in cases where the initially
requested hypotheses do not hold true obtaining correct results.\\
Our approach, differently from all other approaches,  is directly 
founded to the definition of the stress tensor as functional derivative
 of the effective action with respect to the background metric.
All proofs contained in this paper are substantially based on that 
direct definition. \\
We have seen that, although  it is not possible 
performing the analytic continuations involved in the method in all concrete 
cases (this is the same drawback of the $\zeta$ function regularization of 
the effective action), the method is well managed on a theoretical ground.
Indeed, within  our approach, the proof of the conservation
of the stress tensor, the conformal anomaly formula, several
thermodynamical identities are actually very easy to carry out.
 The infinite renormalization is made harmless
by an automatic cancellation  and the finite part is 
clearly highlighted as a residue of a pole of the stress tensor
$\zeta$ function.  It is furthermore clear that the renormalizing terms
are conserved and depend on the geometry locally and thus can be 
thought as parts of geometrical side of the Einstein equations.
Their explicit form can be obtained by the heat kernel expansion as
outlined previously.\\
We have checked the method  considering  several concrete
cases obtaining a perfect agreement with other renormalization procedures.\\
Particular attention has been paied considering the conical manifold,
where some unresolved problems concerning the physical
 interpretation of the obtained results remain when one consider
the conical manifold  as the Euclidean-thermal Rindler space.\\

Concerning the general features of the method presented within this paper,
many ways remain to explore for the future. An important point to study in deep
should be the relation between Wald's axioms  concerning any renormalized
  stress tensor
\cite{bd,fulling,wald} and the stress tensor arising from our approach.
Moreover, the relation between our approach and the usual
point-splitting approach based on  short-distance Hadamard's behavior
of the two-point functions should be investigated.\\
Other possible  generalizations may concern  integer or  half-integer
spinorial fields and gauge theories.

\par \section*{Acknowledgments}
This work has been financially
 supported by the ECT* (European Centre for Theoretical
 Nuclear Physics and Related Areas).\\
I would like to thank R. Balbinot, D.V. Fursaev, D. Iellici, L. Vanzo
and A. I. Zelnikov for valuable discussions  and suggestions.
I am grateful to the Department of Physics of  University of the Trento
for the hospitality during a part of the  time spent to produce this paper

\appendix

\section*{Main formulas}

Let us consider an Euclidean N-manifold ${\cal M}$. Suppose that 
${\cal M}$ is closed (namely compact without boundary. \\
Let $A$ be 
a second-order
 elliptic (positive-definite) selfadjoint  differential operator working
on smooth real  scalar fields of 
$L^2({\cal M}, d\mu_g)$, $\mu_g$ being the usual 
Riemannian measure induced by the Euclidean metric.\\
Let us finally suppose that  the spectrum of the operator 
is discrete. This holds, for example,  in the case of
the Laplace-Beltrami operator with the sign changed, namely the 
$0$-forms Hodge-de Rham Laplacian; in such a case the multiplicity is also 
finite.\\
All that we go to describe should  be more o less 
generalizable  by relaxing some of the 
conditions above, employing opportune spectral measures and so on. 
In particular one could consider the operator $A$ working on
n-forms and deal with the 
Hodge-de Rham formalism also in manifolds non compact
or with boundary. Anyhow,  
this latter case could be more complicated to deal with.
We leave to the mathematicians all these considerations.\\

Our goal is to determine how the generic eigenvalue $\lambda_n$ changes
due to local changes of the metric $g_{ab}$ of the manifolds
keeping fixed the topology.

Let us introduce the Euclidean action
\begin{eqnarray}
S_A[\phi,\phi] := S_A[\phi]
:= -\frac{1}{2} \int_{\cal M} d^N x \:\sqrt{g(x)}\: \phi(x) A\phi(x).
\label{action2}
\end{eqnarray}
Thus we have
\begin{eqnarray}
\frac{\delta S_A}{ \delta \phi(x)} = -\sqrt{g} A \phi(x). \label{variation}
\end{eqnarray}
Let $\lambda_n$ be the eigenvalue of the normalized eigenvector
$\phi_n$, it holds
\begin{eqnarray}
A\phi_n = \lambda_n \phi_n \:\:\:\:,\:\:\:\: 
\int_{\cal M} d^N x \:\sqrt{g(x)}\: \phi_n^*(x)\phi_n(x) = 1,
\label{normalization} \end{eqnarray}
\begin{eqnarray}
\lambda_n = -2 S[\phi_n^*,\phi_n] \label{central}.
\end{eqnarray}
One may change the metric as $g_{ab}(x) \rightarrow g'_{ab}(x) = g_{ab}(x) +
\delta g_{ab}(x)$. 
Obviously, provided opportune mathematical conditions were satisfyed, we expect
 to find a corresponding variation $\lambda_n \rightarrow \lambda'_n =
 \lambda_n +\delta \lambda_n$. We are interested in evaluating the rate of
the variation of the eigenvalues with respect to the metric. In fact,
we want to compute the functional derivative:
\begin{eqnarray}
\frac{\delta \lambda_n}{\delta g^{ab}(x)}
 = -2\frac{\delta S_A[\phi_n^*,\phi_n]}{\delta g^{ab}(x)},
\end{eqnarray} 
where we employed (\ref{central}).\\
Starting from the identity just written above, we have
\begin{eqnarray}
-\frac{\delta \lambda_n}{\delta g^{ab}(x)} = 2 
\int d^N y\: \frac{\delta S_A}{\delta \phi_n^*(y)} 
\frac{\delta \phi^*_n(y)}{\delta g^{ab}(x)} +
2\int d^N y\: \frac{\delta S_A}{\delta \phi_n(y)} 
\frac{\delta \phi_n(y)}{\delta g^{ab}(x)} +
2\frac{\delta_g S_A}{\delta g^{ab}(x)}  \label{f}.
\end{eqnarray}
Using the formula corresponding to  Eq.(\ref{variation}) 
for $\phi_n$ and $\phi_n^*$ (notice that a further factor $1/2$
appears in this case), we obtain:
\begin{eqnarray}
-\frac{\delta \lambda_n}{\delta g^{ab}(x)} = 
-\lambda_n \int d^N y\:\sqrt{g(y)}\: \left(
 \phi_n \frac{\delta \phi^*_n(y)}{\delta g^{ab}(x)} +
\phi^*_n \frac{\delta \phi_n(y)}{\delta g^{ab}(x)}\right) +
2\frac{\delta_g S_A}{\delta g^{ab}(x)}.
\end{eqnarray}
Let us look at the first term in the right hand side of the equation above.
We can rewrite down that as
\begin{eqnarray}
-\lambda_n \int d^N y\:\sqrt{g(y)} \:
\frac{\delta \:\:\:}{\delta g^{ab}(x)} 
\left(\phi_n(y) \phi^*_n(y) \right) = \nonumber
\end{eqnarray}
\begin{eqnarray}
-\lambda_n 
\frac{\delta \:\:\:}{\delta g^{ab}(x)}  \int d^N y\:\sqrt{g(y)} \:
 \phi_n(y) \phi^*_n(y) 
+\lambda_n \int d^N y \:
 \frac{\delta \sqrt{g(y)}}{\delta g^{ab}(x)}  
 \phi_n(y) \phi^*_n(y).  
\end{eqnarray}
The first term in the last line vanishes due to the normalization condition
in Eq.(\ref{normalization}) which is supposed to hold during the variational
process. Eventually, a  few of elementary calculations produces
a well-known result:
\begin{eqnarray}
\frac{\delta \sqrt{g(y)}}{\delta g^{ab}(x)} =
\frac{\partial \sqrt{g(x)}}{\partial g^{ab}(x)} \delta(x-y)
= - \frac{1}{2} \sqrt{g(x)} \: g_{ab}(x) \:\delta(x-y). \nonumber
\end{eqnarray} 
Coming back to the variational derivative of $\lambda_n$ with respect to
the metric and making use of the obtained results in (\ref{f})
 we get our main equation (\ref{start3})
\begin{eqnarray}
\frac{\delta \lambda_n}{\delta g^{ab}(x)} = 
\frac{\lambda_n}{2}\: \sqrt{g(x)} \: g_{ab}(x) \: \phi_n(x)\phi_n^*(x)
- 2\frac{\delta_g S_A[\phi_n^*,\phi_n]}{\delta g^{ab}(x)} \nonumber.
\end{eqnarray}\\
We finally remark that in \cite{rovelli} a similar relation has been found 
in a different context  
as far as eigenvalues of Dirac's operator is concerned.\\

Let us finally prove (\ref{endend}). We  suppose that our closed
manifold is {\em stationary}, namely a global coordinate system exists in
where the Euclidean  metric looks like
\begin{eqnarray}
ds^2 = g_{00}(\vec{x})\:dx^0\:dx^0 + 2g_{0i}(\vec{x})\: dx^0\:d x^{i}
+g_{ij}(\vec{x})\:dx^{i}\:dx^{j} \label{metric}.
\end{eqnarray}
where $\vec{x}\equiv x^{i} \in \Sigma$.
 Notice that $\partial_0$ is a Killing vector.\\
We  suppose also that the manifold (the metric) 
is periodic in the coordinate $x^0$ with a period $\beta$. \\
Our action reads, in the considered coordinates
\begin{eqnarray}
S[\phi] := \int_0^\beta \: dx^0\: \int_{\Sigma}\:d\vec{x}
\sqrt{g(\vec{x})}\: \phi(x) A \phi(x) \nonumber.\:
\end{eqnarray}
Because it will be very useful shortly,
we can consider the new coordinate set given by  $y^0 := x^0/\beta$,
$\vec{y} := \vec{x}$. In those coordinates, posing
$\psi(y) :=\phi(x)$ the action reads:
\begin{eqnarray}
S[\psi] := \int_0^1 \: dy^0\: \int_{\Sigma}\:
\sqrt{f(\vec{y})}\: \psi(y) B \psi(y) \label{actionb}.
\end{eqnarray}
where $B$ is obviously defined with respect to the metric $f_{ab}(y)$ which
reads
$f_{00}(y) := g_{00}(x)/\beta^2$ and $f_{0i}(y) := g_{0i}(x)/\beta$,
$f_{ij}(y) := g_{ij}(x)$.
Now, we observe that, in (\ref{actionb}), variations of the parameter
 $\beta$ can be thought as variation  of the metric of the manifold,
keeping fixed the topology.\\
As for the previous proof it is convenient starting  with the usual identity
\begin{eqnarray}
\lambda_n = -2 \:S[\psi^*_n,\psi_n,f] \nonumber.
\end{eqnarray}
From that it follows
\begin{eqnarray}
\frac{\partial\lambda_n}{\partial\beta}
= -2 \int d^4y \: 
\left\{  \frac{\delta S}{\delta f^{ab}(y)} \frac{\partial f^{ab}}{\partial\beta} 
+ \frac{\delta S}{\delta \psi^*_n(y)}\frac{\partial\psi_n^*}{\partial\beta} +
\frac{\delta S}{\delta \psi_n(y)}\frac{\partial\psi_n}{\partial\beta}
\right\} = \nonumber
\end{eqnarray}
\begin{eqnarray}
=  -2 \int d^4y \: \sqrt{f(y)}
\left\{ 
\frac{-2}{\beta^3} g(y)^{-1/2}
 \frac{\delta S}{\delta f^{00}(y)} f^{00}(y) \beta^2+
\frac{-2}{\beta^2} g(y)^{-1/2}
 \frac{\delta S}{\delta f^{0i}(y)} f^{0i}(y) \beta\right\}+ \nonumber 
\end{eqnarray}
\begin{eqnarray}
  -2 \int d^4y \: \sqrt{f(y)}
\left\{ 
\frac{\partial\psi_n^*}{\partial\beta} 
- \frac{\lambda_n}{2} \psi_n(y)\frac{\partial\psi_n^*}{\partial\beta}
- \frac{\lambda_n}{2} \psi^*_n(y)\frac{\partial\psi_n}{\partial\beta}
\right\} = \nonumber
\end{eqnarray}
\begin{eqnarray}
=-\frac{2}{\beta} \int d^4y \: \sqrt{f(y)}
\: \bar{T}^0_0[\psi^*_n\psi_n](y) + \lambda_n \int d^4y \: \sqrt{f(y)}
\frac{\partial \psi^*_n(y)\psi_n(y)}{\partial\beta} \label{step}.
\end{eqnarray}
Above,
$\bar{T}_{ab}(y)$ is the stress tensor evaluated in the coordinate $y^{a}$.\\
Let us consider the second term in (\ref{step}). We can also write that as
\begin{eqnarray}
\lambda_n \int d^4y \: 
\frac{\partial \sqrt{f(y)} \psi^*_n(y)\psi_n(y)}{\partial\beta}-
\lambda_n \int d^4y \:
\frac{\partial \sqrt{f(y)} }{\partial\beta} \psi^*_n(y)\psi_n(y) =\nonumber
\end{eqnarray}
\begin{eqnarray}
\lambda_n \frac{\partial\:\:}{\partial\beta}\int d^4y \: 
 \sqrt{f(y)} \psi^*_n(y)\psi_n(y)-
\frac{\lambda_n}{\beta} \int d^4y \:
\sqrt{f(y)} \psi^*_n(y)\psi_n(y)\nonumber
\end{eqnarray}
The first term in the right hand side of the equation above
 vanishes due to the invariant normalization condition of
the modes. The second term, as well as the remaining term in (\ref{step}), 
can be translated into the initial coordinates
obtaining
\begin{eqnarray}
\frac{\partial \lambda_n}{\partial\beta} = -\frac{2}{\beta} \int d^4x \: \sqrt{g(\vec{x})}
\: T^0_0[\phi^*_n\phi_n](\vec{x})
+\frac{\lambda_n}{\beta} \int d^4x \:
\sqrt{g(\vec{x})} \phi^*_n(x)\phi_n(x) \nonumber.
\end{eqnarray}
Notice that, as we said above,
 both the integrands do not depend on $x^0$ because the metric
is stationary,
and thus the integration on the temporal variable produces only a factor
 $\beta$.
The final formula is then (\ref{endend}):
\begin{eqnarray}
\frac{\partial \lambda_n}{\partial\beta} = -2 \int_\Sigma d\vec{x}
\sqrt{g(\vec{x})}\: \left\{ T_0^0[\phi^*_n\phi_n](\vec{x}) 
+ \frac{1}{2}\:g^0_0 \:\lambda_n \: \phi_n^*(\vec{x}) \phi_n(\vec{x}) \right\}
\nonumber
\end{eqnarray}

\end{document}